\documentclass{aa}  
\makeatletter
\renewcommand*\aa@pageof{, page \thepage{} of \pageref*{LastPage}}
\usepackage{natbib}
\bibpunct{(}{)}{;}{a}{}{,}
\def\bibfont{\aa@bibliographyfont}
\makeatother
\usepackage[T1]{fontenc}
\usepackage{amsmath}

\usepackage{gensymb}
\usepackage{bm}	
\usepackage{graphicx}
\usepackage{newtxtext}
\usepackage[smallerops,cmbraces]{newtxmath}
\usepackage{xcolor}
\usepackage{colortbl}
\usepackage{nicematrix}
\definecolor{xlinkcolor}{cmyk}{1,1,0,0}
\usepackage[
  bookmarks=true, 
  pdfnewwindow=true,
  colorlinks=true, 
  linkcolor=xlinkcolor,
  citecolor=xlinkcolor, 
  filecolor=xlinkcolor, 
  urlcolor=xlinkcolor, 
  final=true,
]{hyperref}
\usepackage{soul}
\usepackage{xspace} \xspaceaddexceptions{]\}}
\usepackage[normalem]{ulem}
\usepackage{bm}
\usepackage{textgreek}
\usepackage[capitalize]{cleveref}
\crefname{section}{Sect.}{Sects.}

\creflabelformat{equation}{#2#1#3}
\crefname{enumi}{item}{items}

\usepackage{siunitx}
\DeclareSIUnit[number-unit-product = ]\percent{\char`\%}
\usepackage{csquotes}
\newcommand{\txt}[1]{\mathrm{#1}}
\usepackage[nomessages]{fp}
\usepackage{fp-basic}
\usepackage{etoolbox}
\usepackage{pgf}
\definecolor{high}{HTML}{0084a1}
\definecolor{low}{HTML}{bfe0e8}
\definecolor{lightblue}{rgb}{0.1,0.5,0.89}
\newcommand*{\opacity}{90}
\newcommand*{\minval}{0}
\newcommand*{\maxval}{1037}
\newcommand{\gradient}[1]{
    \ifdimcomp{#1pt}{>}{\maxval pt}{#1}{
        \ifdimcomp{#1pt}{<}{\minval pt}{#1}{
            \pgfmathparse{int(round(100*(#1/(\maxval-\minval))-(\minval*(100/(\maxval-\minval)))))}
            \xdef\tempa{\pgfmathresult}
            \cellcolor{high!\tempa!low!\opacity} #1
    }}
}

\begin{document}
\def\textfraction{0}
\def\topfraction{1}
\def\bottomfraction{1}
\def\dbltopfraction{1}
\def\floatpagefraction{.9}
\def\dblfloatpagefraction{.9}
\def\labelitemi{$\bullet$}
\definecolor{blackberry}{HTML}{8D1D75}
\definecolor{lightblue}{rgb}{0.1,0.5,0.89}
\definecolor{teal}{HTML}{20B2AA}
\definecolor{mblue}{rgb}{0.07, 0.50, 0.99}
\definecolor{pinky}{rgb}{0.8,0.04,0.8}
\definecolor{mr}{rgb}{0.95,0.87,0.67}
\definecolor{hr}{rgb}{0.92,0.69,0.94}
\definecolor{uhr}{rgb}{0.52,0.81,098}
\definecolor{xhr}{rgb}{0.85,0.85,0.83}
\definecolor{uhrl}{rgb}{0.65,0.91,0.90}
\newcommand{\magneticum}{\textsc{Magneticum}\xspace}
\newcommand{\magpath}{\textsc{Magneticum Pathfinder}\xspace}
\newcommand{\Bz}{\textsc{Box0}\xspace}
\newcommand{\Bo}{\textsc{Box1a}\xspace}
\newcommand{\Bt}{\textsc{Box2}\xspace}
\newcommand{\Btb}{\textsc{Box2b}\xspace}
\newcommand{\Bth}{\textsc{Box3}\xspace}
\newcommand{\Bf}{\textsc{Box4}\xspace}
\newcommand{\Bfi}{\textsc{Box5}\xspace}
\newcommand{\subfind}{\textsc{SUBFIND}\xspace}
\newcommand{\Msun}{M_\odot}
\def\H4{$H_4$\xspace}
\def\h4{$\langle h_4\rangle$\xspace}
\def\rh{$r_{1/2}$}
\def\th{$\theta$\xspace}
\def\lr{$\lambda_{R_{1/2}}$\xspace}
\def\bH4{$\bm{H_4}$\xspace}
\def\bh4{$\bm{\langle h_4\rangle}$\xspace}
\def\ion#1#2{#1\,\textsc{#2}\xspace}
\def\HI{\ion{H}{i}}
\def\MgII{\ion{Mg}{ii}}
\def\CII{\ion{C}{ii}}
\def\CIV{\ion{C}{iv}}
\def\NII{\ion{N}{ii}}
\def\OII{\ion{O}{ii}}
\def\OIII{\ion{O}{iii}}
\renewcommand\labelitemi{-}

\title{Die Hard: the on-off-cycle of galaxies on the\\ star formation main sequence}
\titlerunning{The on-off-cycle of galaxies}

\author{
    Silvio Fortun\'e\thanks{sfortune@usm.lmu.de}\inst{\ref{inst:origins},\ref{inst:usm}}
    \and
    Rhea-Silvia Remus\inst{\ref{inst:usm}}
    \and
    Lucas C. Kimmig\inst{\ref{inst:usm}}
    \and
    Andreas Burkert\inst{\ref{inst:origins},\ref{inst:usm},\ref{inst:mpe}}
    \and
    Klaus Dolag\inst{\ref{inst:usm},\ref{inst:mpa}}
    }
\authorrunning{Fortuné et al.}

\institute{
    Excellence Cluster ORIGINS, Boltzmannstr.~2, 85748 Garching bei M\"unchen, Germany\label{inst:origins}
    \and
    Universit\"ats-Sternwarte, Fakult\"at f\"ur Physik, Ludwig-Maximilians-Universit\"at München, Scheinerstr.~1, 81679 M\"unchen, Germany\label{inst:usm}
    \and
    Max-Planck-Institut f\"ur Astrophysik, Karl-Schwarzschild-Str.~1, 85748 Garching, Germany\label{inst:mpa}
    \and
    Max-Planck-Institute for Extraterrestrial Physics, Giessenbacherstr. 1, 85748 Garching, Germany\label{inst:mpe}
    }

\date{Received XXX / Accepted YYY}

\abstract
{Our picture of galaxy evolution currently assumes that galaxies spend their life on the star formation main sequence until they may eventually be quenched.
However, recent observations show indications that the full picture might be more complicated.}
{We reveal how the star formation rates of galaxies evolve, possible causes and imprints of different evolution scenarios on galactic features.}
{We follow the evolution of central galaxies in the highest-resolution box of the \magpath cosmological hydrodynamical simulations and classify their evolution scenarios with respect to the star formation main sequence.}
{We find that a major fraction of the galaxies undergoes long-term cycles of quenching and rejuvenation on gigayear timescales. This expands the framework of galaxy evolution from a secular evolution to a sequence of multiple active and passive phases. 
Only $14\%$ of field galaxies on the star formation main sequence at $z \approx 0$ actually evolved along the scaling relation, while the bulk of star forming galaxies in the local universe have undergone cycles of quenching and rejuvenation.
In this work we describe the statistics of these galaxy evolution modes and how this impacts their mean stellar masses, ages and metallicities today.
We further explore possible explanations and find that the geometry of gas accretion at the halo outskirts shows a strong correlation with the star formation rate evolution, while the density parameter as a tracer of environment shows no significant correlation.
A derivation of star formation rates from gas accretion with simple assumptions only works reasonably well in the high-redshift universe where accreted gas gets quickly converted into stars.}
{We conclude that an evolution scenario consistently on the main sequence is the exception, when regarding galaxies on the main sequence at lower redshifts. Galaxies with rejuvenation cycles can be distinguished well from main-sequence-evolved galaxies, both in their halo accretion modes and in their features at $z\approx 0$.}


\maketitle

\section{Introduction}\label{sec:introduction}

The transformation of diffuse gas into stars is a process that spans from large scale structure formation \citep[e.g.][]{vandeweygaert:2008:bond} down to giant molecular clouds that host star forming clumps \citep[e.g.][]{mckee:2007:ostriker}.
As baryonic physics is subject to a chaotic interplay of gravitation and hydrodynamics with cooling, heating and dust formation as well as feedback from stars, supernovae and active galactic nuclei (AGN), it is remarkable that structure and star formation in galaxies produces characteristic relations and evolutionary pathways.
In a color-magnitude diagram, active galaxies form a “blue cloud” separated from the “red sequence” \citep[e.g.][]{strateva:2001,balogh:2004} with fewer galaxies occupying the “green valley” in between \citep[][]{wyder:2007}.
Additionally, active galaxies are observed to follow a relation between their stellar masses and star formation rates, which is called the “star formation main sequence” (SFMS) \citep[e.g.][]{speagle:2014,pearson:2018,leslie:2020,popesso:2023}.
Consistent findings of a declining star formation main sequence toward lower redshifts show that the cosmological conditions for conversion of gas into stars change with redshift \citep[e.g.][]{teklu:2023}.
This can also be seen from studies of the star formation rate density (SFRD), which evolves with a universal peak at $z \approx 2$ and lower total SFRD at both higher and lower redshifts \citep[][]{madau:2014:dickinson}.
Their results for the SFRD evolution are consistent with a general decline of specific star formation rates ($\txt{sSFR} = \txt{SFR}/\txt{M}_*$) in galaxies at later times, which corresponds with a decline of the overall star formation main sequence.
However, the different mass- and redshift-dependent descriptions of the main sequence \citep[e.g.][]{pearson:2018, santini:2017, speagle:2014} show that the physics behind the scaling relation are not yet fully understood.

\citet{lin:2024} used ALMA and MaNGA observations and found “green-valley” galaxies that displayed low star-formation rates despite having similar amounts of dense gas as main-sequence galaxies.
Even with constant and smooth accretion, internal processes of a galaxy are expected to lead to a scatter of $\Delta\txt{MS} = \txt{log}(\txt{SFR}/\txt{SFR_\txt{MS}}) \approx 0.3$ dex around the main sequence at different redshifts \citep[e.g.][]{noeske:2007a,whitaker:2012,speagle:2014,pearson:2018}.
\citet{tacchella:2016} tested the scatter at $z>1$ using results from the \textsc{Vela} zoom-in simulation suite and attribute this to a cycle of compaction, depletion and accretion in galaxies until gas inflow from the cosmic web comes to an end.
Mergers provide an additional mode of increased star formation activity, either as a consequence of a wet-merger-induced starburst or through enhanced gas compaction via interaction with a galaxy \citep[][]{gomez-guijarro:2022}.
In our current picture of galaxy formation, it is usually assumed that galaxies evolve along the main sequence until they are quenched by one of the commonly discussed processes, such as galaxy mergers \citep[e.g.][]{kormendy:2013:ho,ellison:2022} or halo mass quenching \citep[][]{birnboim:2003:dekel,dekel:2008:birnboim}.
However, cases of rejuvenation have been discussed lately, as evidence for this is seen at high redshifts \citep[][]{remus:2023, trussler:2025} but also at low redshifts \citep[e.g.][]{yi:2005,chauke:2019}.
Recently, models also predict that rejuvenation might be a common occurrence in galaxy formation \citep[e.g.][]{zhang:2023,tanaka:2024}.
However, as of yet it is unclear how common this process really is and how often a galaxy simply evolves along the main sequence, as commonly assumed.

The principle of mass conservation in its application as the “Bathtub” gas regulator model for smooth accretion \citep[e.g.][]{bouche:2010,lilly:2013,burkert:2017} provides a simple framework to test our understanding of the interplay of gas flows and star formation.
This interplay is governed by the depletion time $t_\txt{depl} \equiv M_\mathrm{gas}/\mathrm{SFR} \sim 1\ \txt{Gyr}$ \citep[][]{tacconi:2020}, where processes like recycling of gas in a “halo fountain” \citep[][]{oppenheimer:2008:dave,tumlinson:2017} and cooling become important.
In order to identify typical star formation histories and quantify their prevalence, we study and classify long-term trends of galaxies in the \magpath cosmological simulations\footnote{www.magneticum.org}.
This paper is structured as follows:
In \autoref{sec:magneticum}, we describe the simulation used for this work.
Then, we describe in \autoref{sec:classes} the method employed to identify moments of quenching and rejuvenation as well as how we use them to classify the formation histories of the galaxies in our sample.
We then show how these classes differ at $z\approx 0$ with respect to their mean ages, metallicities and mass distributions in \autoref{sec:imprints}.
Finally, we investigate causes of the different characteristic formation classes by measuring gas flow rates, gas flow geometries and environment in sections ~\ref{sec:accretionrates} and ~\ref{sec:environment}.

\section{The \magneticum simulations}\label{sec:magneticum}

We perform our analysis using Box4~uhr of the \magpath cosmological hydrodynamical simulation suite \citep{teklu:2015}. 
These simulations are based on the cosmological parameters from the WMAP7 results according to \citet{komatsu:2011} with a mass density of $\Omega_\txt{m} = 0.272$, a baryon fraction of $f_\txt{baryon} = 0.168$, a cosmological constant of $\Lambda = 0.728$ and a dimensionless Hubble constant of $h = 0.704$.
The simulations were run with a modified version of the smoothed-particle hydrodynamics (SPH) code GADGET-2 \citep[][]{springel:2001b,springel:2005a} with updates regarding the SPH kernel and treatment of viscosity \citep[][]{dolag:2005,beck:2016,donnert:2013}.

The baryonic physics includes gas cooling, star formation and stellar winds as described by \citet{springel:2003:hernquist}. 
Accordingly, gas particles include contributions to the hot and cold phases with star formation based on the latter.
Exchange between the two phases is implemented through radiative cooling on one hand with rates based on cooling tables from \citet{wiersma:2009}.
Their results are based on calculations using the code \textsc{cloudy} \citep[][]{ferland:1998} while considering contributions from the eleven elements H, He, C, N, O, Ne, Mg, Si, S, Ca and Fe, which are traced individually within the simulation for all stellar and gas particles, with exposition to the cosmic microwave background radiation and the UV/X-ray background radiation from galaxies and quasars according to the model by \citet{haardt:2001:madau}.

Star formation, supernovae (SNe) and chemical enrichment are implemented as a subresolution model according to \citet[][]{tornatore:2003,tornatore:2007,dolag:2017} with initial stellar population distribution according to \citet{chabrier:2003}. 
The SN feedback is implemented partly as heating and partly as galactic winds with $v_\txt{wind} = 350$ km/s triggered from a 10\%-fraction of SNe II ($10^{51}$ erg). 
Black hole growth and AGN feedback are modeled as described by \citet[][]{springel:2005b,dimatteo:2005} with modifications covered by \citet{fabjan:2010} and \citet{hirschmann:2014}.
AGN luminosities are derived from the BH accretion rate as $L_\txt{r}= \epsilon_\txt{r}\ \Dot{M}_\bullet\ c^2$, with a radiative efficiency of $\epsilon_\txt{r} = 0.1$ for a non-rapidly spinning black hole \citep[][]{shakura:1973:sunyaev}.
Of the total radiated energy, $\epsilon_\txt{f}=10\%$ is thermally coupled to the surrounding gas, leading to a feedback energy rate of $\Dot{E}_\txt{f} = \epsilon_\txt{r}\ \epsilon_\txt{f}\ \Dot{M}_\bullet\ c^2$.
Thermal conduction in galaxy clusters is implemented as described by \citet{dolag:2004} but with a value of $1/20$ (instead of $1/3$) of the Spitzer value \citep[][]{spitzer:1962}, since anisotropic conduction effects create a net heat transport corresponding to a Spitzer value of one percent, while values of more than ten percent lead to oversmoothed temperature profiles, as shown by \citet{arth:2014}.

With a box size of $48\ \txt{h}^{-1}\txt{Mpc}$, the simulation volume contains 1318~central galaxies with a final $z=0$ stellar mass of $M_* \geq 1\times 10^{10} M_\odot$.
These central galaxies are the \subfind-identified dominant galaxies of each halo lying at the center of the potential well, thus excluding satellites. For the given box size, our sample represents primarily field galaxies, with only a few group and cluster centrals.
We choose central galaxies with this stellar mass threshold in order to reliably trace galaxies, including their kinematics, with a sufficient particle resolution of $N_*>5000$.
This is a consequence of the resolution given by particles masses of $m_\txt{dm} = 3.6\times 10^7 \txt{h}^{-1}M_\odot$ and $m_\txt{gas} = 7.3\times 10^6 \txt{h}^{-1}M_\odot$ and softening lengths of $\epsilon_\txt{dm/gas} = 1.4 \txt{h}^{-1}\txt{kpc}$ and $\epsilon_\txt{*} = 0.7 \txt{h}^{-1}\txt{kpc}$.
Each gas particle can spawn a star particle up to four times resulting in lower masses for stellar particles of $\txt{m}_* \approx 1.8\cdot 10^6 \txt{h}^{-1}M_\odot$.
Gas particles are not fully consumed upon forming a star particle but instead are capable of spawning up to four star particles, which is a unique advantage of the \magpath simulations. Not only does this increase the stellar mass resolution by a factor of four, but it also better accounts for the more typical process of giant molecular clouds to partially outlive star formation and absorb momentum and chemical enrichment from stellar feedback \citep[e.g.][]{chevance:2023}. Specifically, a gas particle can form stars at its current metallicity, then be further enriched and ejected by the stellar feedback and potentially return to resume star formation later.

The simulation has been shown to reproduce several critical galaxy properties in previous works, for example with respect to scaling relations across different redshifts \citep[e.g.][]{teklu:2015,teklu:2017,remus:2022, dolag:2025}, the kinematical properties of galaxies \citep[e.g.,][]{schulze:2018,vdsande:2019,valenzuela:2024}, and the star formation properties of galaxies \citep[e.g.][]{kudritzki:2021,teklu:2023}. It is especially noteworthy that the galactic scaling relations arise naturally while the tuning of the \magpath simulations was towards hot halos of galaxy clusters. Furthermore, the simulation suite captures galaxy properties at the higher redshifts of $z \approx 4$, which is of particular difficulty \citep[e.g.][]{lustig:2023,remus:2023,kimmig:2025a,dolag:2025}, but especially relevant in the context of this study where we trace galaxy properties through cosmic time.

\begin{figure*}[h!]
  \begin{center}
    \includegraphics[width=0.9\textwidth]{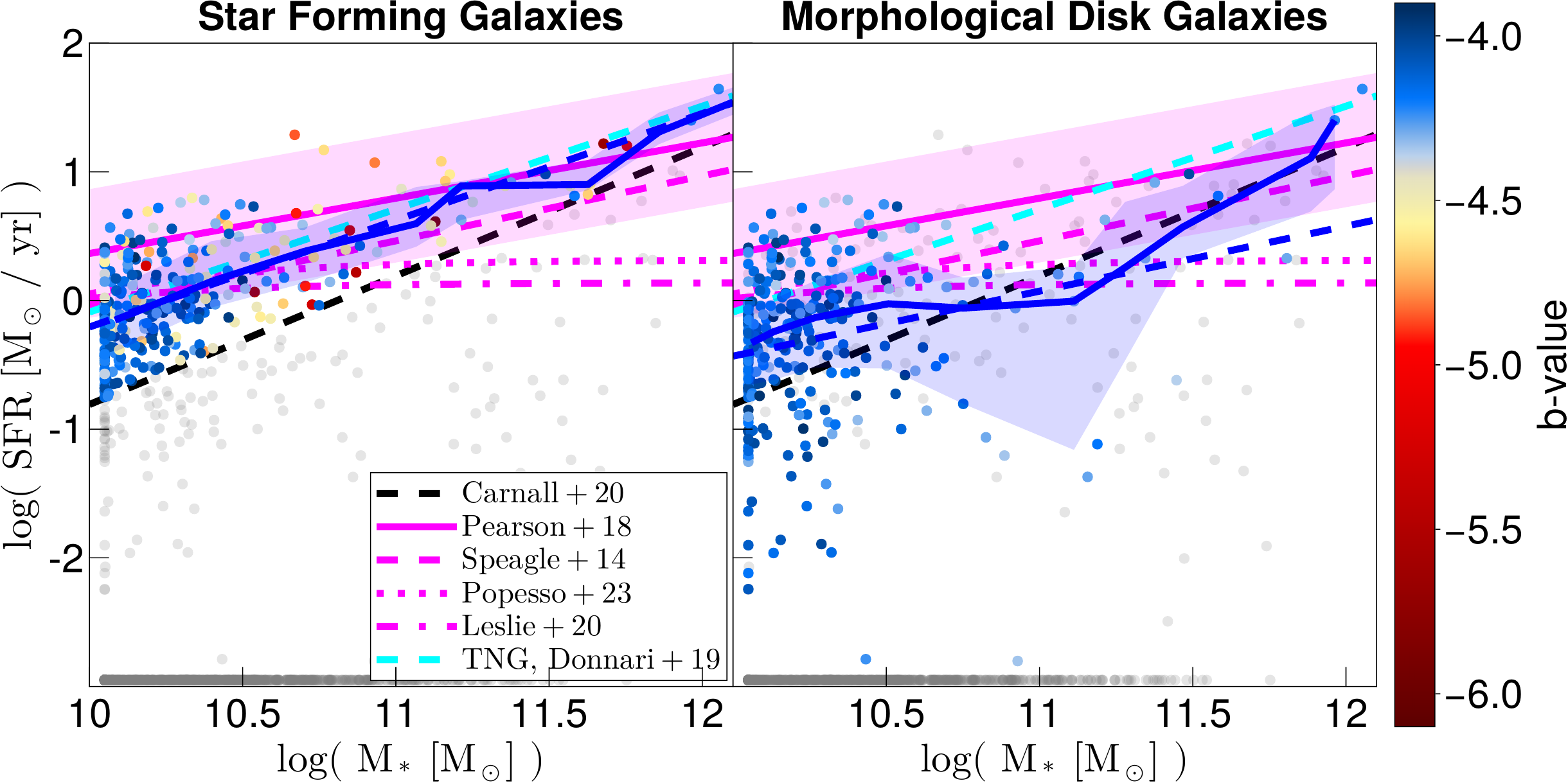}
    \caption{
    The star formation main sequence at $z \approx 0$.
    The left- and right-hand panels show the results with respect to two selection criteria for \magneticum galaxies.
    By imposing a threshold of $\mathrm{sSFR} > 0.2 / t_\mathrm{Hub}$ (left) as used by \citet{carnall:2020} (dashed black line), we find a slightly higher linear fit to the logarithmic values (dashed blue line) compared to selecting for all kinematic disks with $\txt{SFR}>0\ \txt{M}_\odot / \txt{yr}$ (right).
    Results from observations are displayed in magenta, simulations are shown in blue colors.
    The solid blue line and shaded region show the running median and 50th-percentile region.
    The circles represent the galaxies from the \magpath simulations colored according to their morphology as defined by the b-value \citep[][]{teklu:2015}.
    Gray circles are ignored for the running median lines on each panel.
    SFR values below the range of the figure are fixed to the bottom to represent the full sample in this work.
    The magenta solid line and shaded region highlights a range of $\pm 0.5\ \txt{dex}$ around the main sequence fit by \citet{pearson:2018}, which we used in this work to distinguish between on and off the main sequence.
    The other magenta lines are additional observational results as dashed \citep[][]{speagle:2014}, dash-dotted \citep[][]{leslie:2020} and dotted \citep[][]{popesso:2023}.
    The dashed cyan line marks the results from \textsc{IllustrisTNG} by \citet{donnari:2019err}.
	}
	\label{fig:mainsequencecomparison}
    \end{center}
\end{figure*}

We traced a set of galaxies by following the most massive progenitors along halo merger trees of \subfind-identified galaxies with a baryonic adaptation by \citet{dolag:2009} of the work by \citet{springel:2001a}.
To clean the sample from rare cases of broken merger trees, we disregard galaxies with an abrupt termination of their merger tree at $z<1$, since this makes it difficult to compare with galaxies that are traced back to cosmic noon and beyond.
Additionally, we remove galaxies that do not enter the star formation main sequence before $z>1$, reducing out final sample from~1318 to 1282~galaxies, which we plot in \autoref{fig:mainsequencecomparison}.

\autoref{tab:classify_references} in \autoref{sec:classification} shows that the latter criterion of a period of star formation on the main sequence impacts the total number of galaxies considered.
This is especially evident when comparing the total number of galaxies classified with respect to the main sequence by \citet{speagle:2014}, for which only 1014~galaxies are among the formation classes discussed in this work.

\section{The main sequence and star formation history classes}\label{sec:classes}

The relation of galaxy stellar mass and star formation rate has long been known to establish a fundamental difference between two groups of galaxies, with those galaxies that are star forming living on a relatively tight star formation main sequence, and those that have quenched their star formation lying within a more diffuse cloud below \citep[e.g.][]{speagle:2014,pearson:2018,popesso:2023,leslie:2020,brinchmann:2004}. This main sequence of galaxies evolves with time, reaching higher star formation rates at a given stellar mass at higher redshifts than at lower redshifts, demonstrating an overall decline in star formation to lower redshifts \citep[e.g.,][]{speagle:2014,pearson:2018}. While this global picture is well established, the observed main sequence relations strongly differ between different surveys already at $z=0$, as can be seen from Fig.~\ref{fig:mainsequencecomparison}, where the different observed relations are shown as pink lines.

We start by testing whether the simulation can reproduce the observed main sequences. Fig.~\ref{fig:mainsequencecomparison} shows in both panels the star formation main sequence found for the \magneticum simulations, in comparison to the results reported from different observations and the IllustrisTNG simulations \citep[][dashed cyan line]{donnari:2019}.
The left panel shows all galaxies colored according to their morphology as traced by the b-value \citep{teklu:2015}, with blue colors indicating disk galaxies and red colors indicating spheroidals.
Yellow colors represent intermediate galaxies. The main sequence from the simulations here is calculated from all galaxies that have specific star formation rates larger than $\mathrm{sSFR} > 0.2 / t_\mathrm{Hub}$, the threshold criterion defined by \citet{carnall:2020}, while galaxies below this threshold are plotted in gray. This criterion is similar to that by \citet{franx:2008} commonly used in the literature to split quiescent from star forming galaxies and included in the figure as black dashed line. The resulting main sequence from \magneticum using this definition is marked as a blue solid line, with the 50th-percentile scatter shown as a blue shaded area.
The main sequence found for the \magneticum simulations using this definition mirrors the results reported for the IllustrisTNG simulations by \citet{donnari:2019}. It well agrees with the observational relations, though we note the existence of some scatter, such that the observations by \citet{pearson:2018} tend to higher SFRs at lower stellar masses compared to \magneticum, while \citet{leslie:2020} and \citet{popesso:2023} find lower SFRs for the more massive galaxies at $z=0$.

The right panel of Fig.~\ref{fig:mainsequencecomparison} shows the main sequence as obtained if only those galaxies that are morphologically classified as disk galaxies are used, with the intermediates and spheroidals plotted in gray. While this criterion does not explicitly select for star-forming galaxies, the absence of blue circles at the bottom of the panel shows that disk galaxies also have higher SFR values. As can be seen immediately, the main sequence obtained from the simulation now is overall at lower values, resembling more the shape of the relation reported by \citet{popesso:2023} or \citet{leslie:2020} before a sudden increase of star formation at $M_* > 10^{11.5} \txt{M}_\odot$. This is due to the existence of what is commonly referred to as red spirals \citep[e.g.][]{masters:2010}, i.e., galaxies that are morphologically spiral galaxies but their star formation has (partially) been quenched. Their presence demonstrates that the morphology of galaxies is not directly correlated to their star formation properties, even though for the large majority there is the trend of disk galaxies being star forming. This was already indicated in the left panel of Fig.~\ref{fig:mainsequencecomparison}, where the coloring of the data points revealed that several of the galaxies on the main sequence according to their star formation properties are in fact of intermediate morphology (i.e., contain for example large bulges) or even, in a few cases, spheroidals, especially at the higher mass end.

Consequently, defining a star formation main sequence is not perfectly straightforward.

\subsection{Formation history classification}\label{sec:classification}
We want to understand if galaxies live on the main sequence until quenching or not.
Thus, we trace galaxies across cosmic time and classify them according to their star formation rate evolution relative to the star formation main sequence by \citet{pearson:2018}. 
While it is evident from Fig.~\ref{fig:mainsequencecomparison} that the relation found by \citet{pearson:2018} does not fit the best to the Magneticum galaxies, we chose to use their best-fit relation as a reference as they provide a consistent method in obtaining the main sequence covering a large range of redshifts, which is required to study the long-term evolution of galaxies. In addition, their values are based on the same cosmology from WMAP7 results \citep[][]{komatsu:2011} as well as the IMF by \citet{chabrier:2003}. However, we do not limit the study to the relation found by \citet{pearson:2018} but rather explore other main sequence definitions as baseline as well, which we will mention throughout this work if differences to the results based on the main sequence by \citet{pearson:2018} are found. Overall, the results presented in this work do not change with the choice of base main sequences, and only the relative fractions of properties vary depending on that choice. An overview of the main sequences used as references and the obtained results is given in Tab.~\ref{tab:classify_references}.
For each snapshot, we define each galaxy as:
\begin{itemize}
    \item quiescent if it is more than 0.5~dex below the main sequence value $\txt{SFR_\txt{MS}(M_*,z)}$, reaching down to SFR$=0$,
    \item star forming if it is within a $\pm 0.5$~dex range of the main sequence value,
    \item and star bursting if the SFR is more than 0.5~dex above the main sequence.
\end{itemize}
Since we look at long-term trends, we focus on the duration of quiescence instead of intensity.
In order to reveal long-term trends, a running median with a window of 1~Gyr is applied first on the star formation rates. This smoothed SFR is compared to the SFR of the main sequence to define the galaxy as quiescent, star forming or star bursting at the given time. This classification is again smoothed over a window of 1~Gyr to identify long-term transitions between quenching and rejuvenation. We require for a change in classification from star forming to quiescent (or vice-versa) that the galaxy remains with the new state for at least $80\%$ of this period of 1~Gyr.

\begin{figure}
  \begin{center}
    \includegraphics[width=1.\columnwidth]{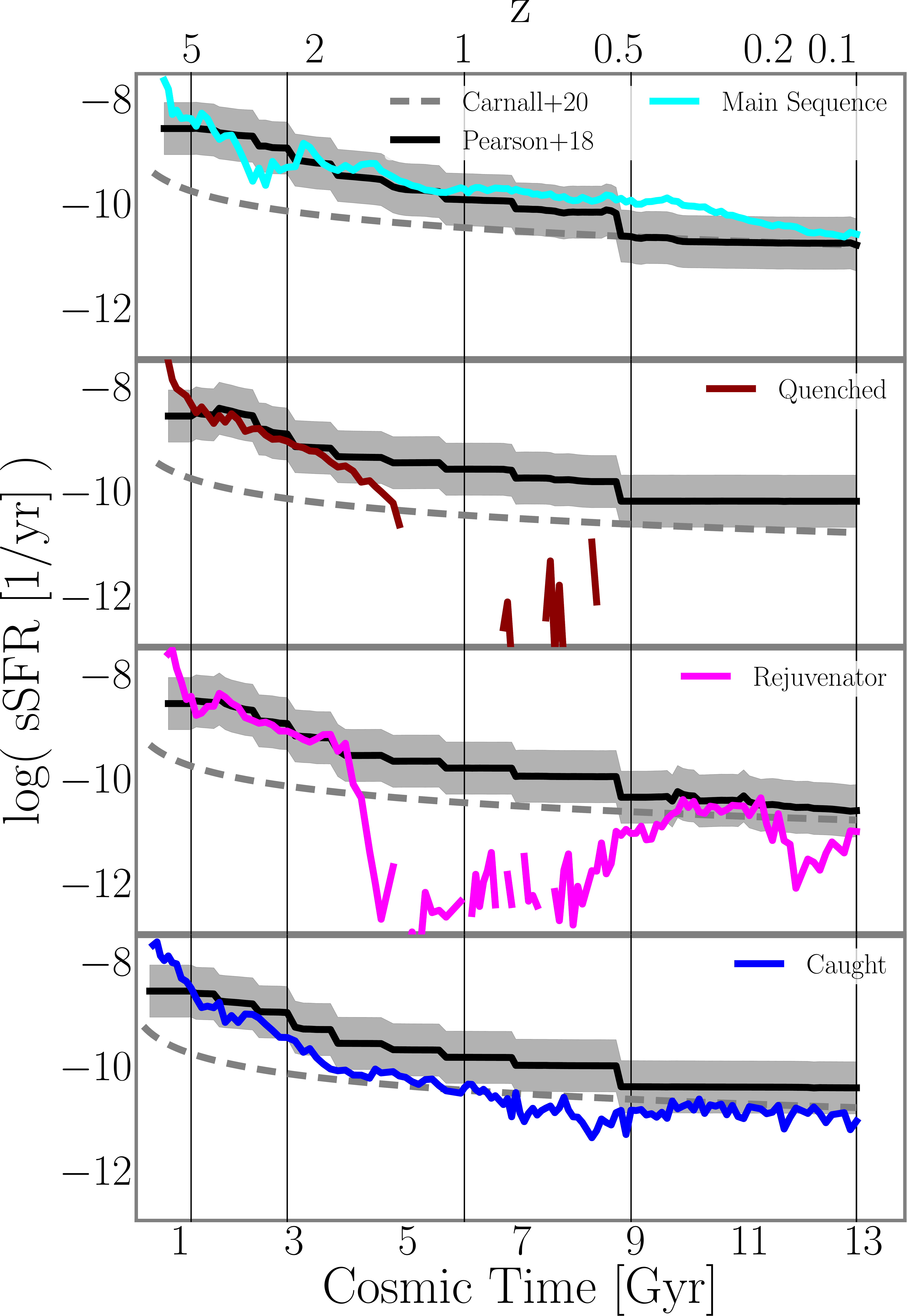}
    \caption{
        Examples of the four formation scenarios for galaxies with respect to specific star formation rate changes over time.
        The black lines and shades indicate the main sequence values depending on the redshift and stellar mass of a galaxy according to results by \citet{pearson:2018} with a scatter of $\pm 0.5\ \txt{dex}$ applied to differentiate between on or off the main sequence.
        The panels show a main-sequence-type galaxy (cyan), an early-quenched galaxy (dark red), a rejuvenating galaxy (magenta) and a caught galaxy (blue) from top to bottom.
        Two classes are not explicitly represented here. 
        First, the “late-quenched” types only differ from the “early-quenched” by dropping off the main sequence after redshift $z=1$. 
        Second, rejuvenator-main-sequence-type galaxies recover a stable position on the main sequence at redshift $z=0$, unlike rejuvenator-quenched-type galaxies which also go through rejuvenation cycles but do not end up on the main sequence. 
        The vertical lines highlight the snapshots corresponding to each row in \autoref{fig:mst_p1e10}.
    }
    {\label{fig:cherrytrack}}
    \end{center}
\end{figure}

Following this approach, we find four different characteristic scenarios of evolution as shown in \autoref{fig:cherrytrack}. We plot a given galaxy's specific star formation rate with cosmic time (colored lines), compared to the expected main sequence value for the galaxy's current stellar mass (black lines and shaded areas), with four representative galaxies for each category (panels going from top to bottom):
\begin{itemize}
    \item Main-sequence galaxies (cyan, top panel) maintain a high star formation rate and never go through periods dominated by quiescence. Only about $2\%$ of the sample never quench.
    \item Quenched galaxies (dark red, second panel) quench once and never recover a SFR within 0.5~dex of the main sequence. Galaxies that quench permanently make up $65\%$ of the sample.
    \item Rejuvenating galaxies (magenta, third panel) quench below the main sequence, but recover a SFR according to the main sequence value. Such rejuvenation cycles are found to occur for individual galaxies up to three times in succession. Rejuvenating galaxies form the second-largest class with $\sim 32\%$ of the sample.
    \item Caught galaxies (blue, bottom panel) are dubbed by virtue of maintaining a constant SFR until the main sequence itself catches it due to its decline with lower redshifts. These formation scenarios were classified manually and only make up $\sim 2\%$ of the sample.
\end{itemize}
Few galaxies remain on or above the main sequence throughout their entire evolution and will be referred to as main-sequence-type galaxies.
Quenched-type galaxies drop and stay below the main sequence until $z \approx 0$.
Rejuvenating galaxies hop on and off the main sequence once or multiple times.
The caught-type class consists of the galaxies that drop below the main sequence but maintain a constant star formation until they passively get “caught” by the main sequence.
This is possible since the main sequence itself decreases with lower redshifts.
Seven galaxies surpass their expected star formation rates for a main sequence galaxy by $0.5$~dex at least 80\% within a 1 Gyr time window.

The study was performed using the star-forming main sequence fit by \citet{pearson:2018}. 
However, we have reproduced our classifications using also other fits from observations as shown in \autoref{tab:classify_references}, as well as the quiescence limits by \citet[][]{franx:2008,carnall:2020}.
While the absolute numbers of galaxies in each class change, a consistent picture emerges, with only a very small fraction of galaxies evolving along the star-forming main sequences.
The bulk of galaxies quenches at high redshift and a consistently significant fraction of galaxies hops on and off the main sequence, of which also a large part end up on the main sequence at $z \approx 0$.
As a result, we find that the dominant group of galaxies on the main sequence at $z \approx 0$ have experienced passive phases on gigayear time scales, independent of the exact main sequence definition used for classification. 
Consequently, our classifications of these different star formation behaviors are robust and not simply definition dependent.
We do note, however, a systematic difference between the main sequence results depending on wether or not a turnover mass was included in their fit. Those main sequence fits which do \citep[][]{leslie:2020,popesso:2023} result in an overall greater number of rejuvenated galaxies (`RM.' in \autoref{tab:classify_references}) and less galaxies that always remained on the main sequence (MS) as compared to those which do not include a turnover mass \citep[][]{speagle:2014,pearson:2018}.
This is expected due to the redshift evolution of each main sequence description. While the relations by \citet[][]{leslie:2020,popesso:2023} require higher star formation rates at high redshift than the relations by \citet{speagle:2014,pearson:2018}, it is the other way round towards $z=0$.
In addition, the turnover mass amplifies this effect for more massive galaxies, leading us to the conclusion that rejuvenation is extremely common among galaxies on main sequences with turnover mass.

\FPeval{\resultpms}{round(34/212,2)}
\FPeval{\resultpc}{round(6/20,2)}
\FPeval{\resultpr}{round(46/123,2)}
\FPeval{\resultpq}{round(256/630,2)}
\FPeval{\resultpms}{round(34/212,2)}
\FPeval{\resultpms}{round(34/212,2)}
\FPeval{\resultpms}{round(34/212,2)}
\FPeval{\resultpms}{round(34/212,2)}
\FPeval{\resultpms}{round(34/212,2)}
\FPeval{\resultpms}{round(34/212,2)}
\FPeval{\resultpms}{round(34/212,2)}
\begin{table}[]
    \caption{Numbers of galaxies within each formation class with respect to expectations according to a list of results for the SFMS.
    }
\centering
\begin{tabular}{|p{0.45cm}|p{1.05cm}|p{0.99cm}|p{0.7cm}|p{1.1cm}|p{0.8cm}|p{0.9cm}|}
    \hline
     & Pearson & Speagle & Leslie & Popesso & Franx & Carnall\\ 
     \hline
    MS.& \gradient{22} & \gradient{24} & \gradient{7} & \gradient{18} & \gradient{1} & \gradient{7} \\
    Q. & \gradient{833} & \gradient{729} & \gradient{707} & \gradient{729} & \gradient{1021} & \gradient{1037} \\
    R. & \gradient{407} & \gradient{245} & \gradient{466} & \gradient{522} & \gradient{114} & \gradient{155} \\
    Ct. & \gradient{20} & \gradient{16} & N.A. & N.A. & N.A. & N.A. \\ \hline \hline
    QE. & \gradient{646} & \gradient{612} & \gradient{678} & \gradient{677} & \gradient{916} & \gradient{826} \\
    QL. & \gradient{187} & \gradient{117} & \gradient{29} & \gradient{52} & \gradient{105} & \gradient{211} \\
    RQ. & \gradient{290} & \gradient{120} & \gradient{215} & \gradient{279} & \gradient{105} & \gradient{132} \\
    RM. & \gradient{117} & \gradient{125} & \gradient{251} & \gradient{243} & \gradient{9} & \gradient{23} \\ \hline
\end{tabular}
    \tablefoot{
        Classes from top to bottom: “main sequence”, “quenched”, “rejuvenating” and “caught”. References from left to right: \citet{pearson:2018,speagle:2014,leslie:2020,popesso:2023}.
        The active-quiescent limits applied in the work by \citet{franx:2008,carnall:2020} were also used for comparison with an approach different from the SFMS.
        Caught-type galaxies were manually identified, which was only done for the first two references listed.
        Subclasses of quenched and rejuvenating galaxies are listed in the four lower rows with “quenched early” (i.e. at $z>1$), “quenched late” (i.e. at $z<1$), “rejuvenating quenched” and “rejuvenating main sequence” galaxies.
        Note that the total number of galaxies varies due to the sample selection criteria as discussed in \autoref{sec:magneticum}.
    }
  {\label{tab:classify_references}}
\end{table}

\begin{figure*}
  \begin{center}
    \includegraphics[width=.8\textwidth]{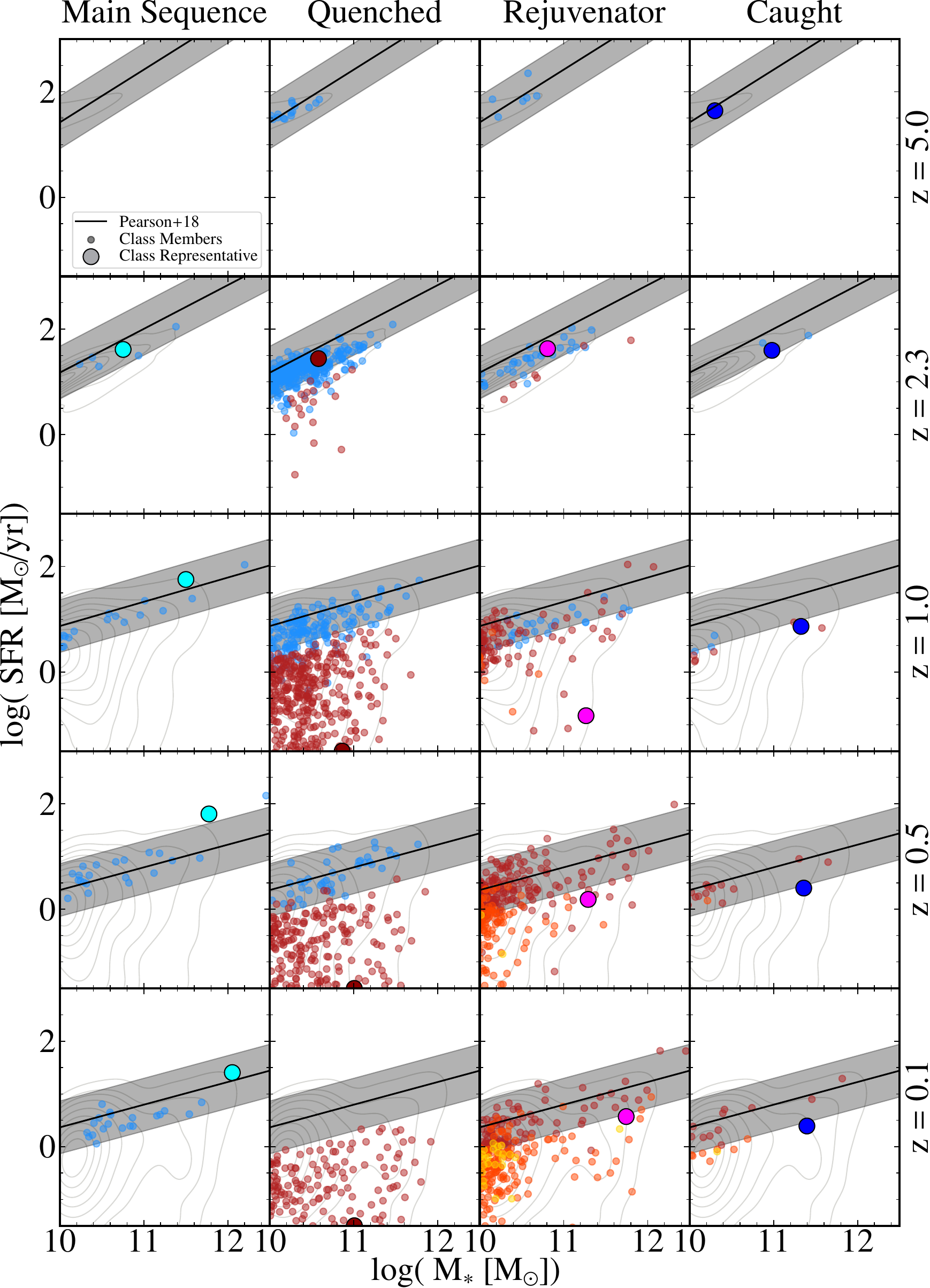}
    \caption{
        Star formation main sequence evolution of the \magneticum galaxies traced back from the selection made at z=0 (bottom-row panels). 
        The rows contain snapshots with decreasing redshift from top to bottom between $6<z<0$. 
        From left to right, the different sub-samples “Main Sequence”, “Quenched”, “Rejuvenator” and “Caught” are highlighted by colors in front of the total sample in gray in the background. 
        Light blue data points represent galaxies that have not yet dropped below the main sequence; each quenching event is counted and encoded in the changing color of the data points as red (once), orange (twice) and yellow (three times).
        The large points are the cherry-picked representative galaxies from \autoref{fig:cherrytrack}.
        Quenched representative galaxies are set as half-circles to the bottom of the panels for visibility.
        The black line and gray shades correspond the main-sequence found by \citet{pearson:2018}. 
        Gray contours outline the distribution of the full galaxy sample for comparison in each panel.
    }
  {\label{fig:mst_p1e10}}
\end{center}
\end{figure*}

In \autoref{fig:mst_p1e10} we plot the star formation rate versus stellar mass for galaxies of each class (columns, going from left to right as main sequence, quenched, rejuvenator and caught) from $z=5$ down to $z=0.1$ (rows, top to bottom).
The circle colors in this figure indicate the number of transitions a galaxy has undergone from star-forming to passive.
A galaxy that has not yet fallen off the main sequence by the respective redshift is shown in light blue. Galaxies that have quenched once appear as red circles, so a red circle on the main sequence has rejuvenated. Orange colors indicate a second transition into quiescence after rejuvenation and yellow colors indicate a third drop off the SFMS after two rejuvenation cycles.
The thin black contour lines show the distribution of our full galaxy sample at each redshift for comparison. We also show the main sequence as defined by \citet{pearson:2018} as a black line (with the shaded area representing $\pm0.5$~dex).

As noted, we find the main bulk of our galaxies to be either quenched or rejuvenators. The former show a large spread in quenching times, with some falling off the sequence as early as $z=2.3$, while others remain on the main sequence all the way down to $z=0.5$ before finally quenching. This is true also for the rejuvenators, except that they continue returning back up to the main sequence. We find that most rejuvenator galaxies undergo their first quenching process at $2>z>1$ just after the cosmic peak of star formation \citep[][]{behroozi:2013,madau:2014:dickinson}.

In the following, we will discuss each of the classes separately in more detail with the main sequence by \citet{pearson:2018} as a reference.

\subsection{Main sequence galaxies}\label{sec:mainsequence}

Only $\sim 2\%$ of all galaxies in our sample and $14\%$ of galaxies on the main sequence at $z\approx 0$ manage to always maintain a high star formation rate. Explicitly, this means that they remain on the star formation main sequence from the moment of first appearing on it until the end of the simulation. While galaxies in this class might briefly drop below 0.5~dex of the main sequence, they quickly recover within less than 1 Gyr and show no lasting periods of quiescence. 

\subsection{Quenched galaxies}\label{sec:quenched}

\begin{figure}
  \begin{center}
    \includegraphics[width=.45\textwidth]{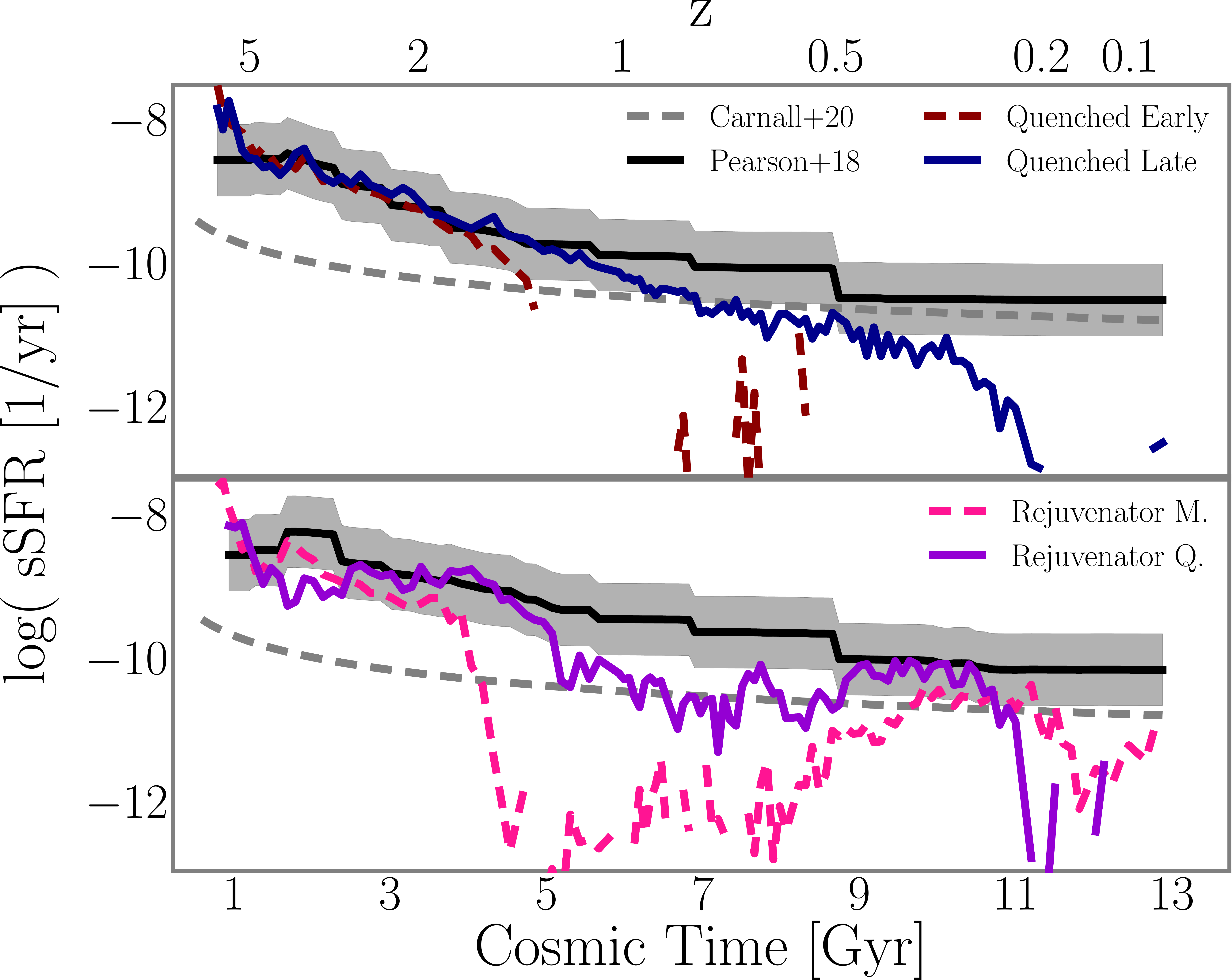}
    \caption{
        Similar to \autoref{fig:cherrytrack} but focused on the difference between the subclass pairs of early- and late-quenched galaxies as well as quenched and main-sequence rejuvenators. 
        The colored dashed lines represent the cases also shown in \autoref{fig:cherrytrack}, while the colored solid lines are the subtypes to differentiate.
        The gray dashed quiescence criteria \citep[][]{carnall:2020}) and black-shaded SFMS expectations \citep[][]{pearson:2018} are derived from the stellar masses of the latter.
        This causes the pink dashed main-sequence-rejuvenator to not lie on the SFMS at $z\approx 0$ due to the larger stellar mass (see \autoref{fig:mssubclasses_1e10}) and therefore lower sSFR.
        The early-quenched galaxy does recover star-formation after quenching but never consistently makes it back onto the main sequence. 
        The late-quenched galaxy also almost makes a passive rejuvenation akin to caught-type galaxies, since the main sequence drops at $z<0.5$. 
        However, it is still mostly more than 0.5~dex below the main sequence averaged over a smoothing window of one gigayear, ruling out a rejuvenation scenario.
        The main-sequence rejuvenator undergoes a more pronounced quenched phase than the quenched rejuvenator.
    }
  {\label{fig:subclass_cherries}}
\end{center}
\end{figure}

About $65\%$ of all galaxies in our sample belong to the quiescent group, remaining on the main sequence until a quenching event brings them below by at least $0.5$~dex, after which they never recover significant amounts of star formation.
Given the difference in quenching times noted in \autoref{fig:mst_p1e10}, the class of quenching galaxies is sub-divided into galaxies that quench early, i.e. before $ z = 1 $ and those that quench late, after $ z = 1 $, well after the peak of star formation. 
We highlight the difference between these two scenarios in the upper panel of \autoref{fig:subclass_cherries}, where we show the specific star formation rate against cosmic time (as in the style of \autoref{fig:cherrytrack}) for two examples of early (red) and late (blue) quenched galaxies.
The early-quenched galaxy drops below the main sequence much more abruptly than the late-quenched galaxy. Interestingly, this example galaxy does rekindle star formation but not consistently and high enough to rejuvenate onto the main sequence. By contrast, the late-quenched galaxy shows a smooth decline in sSFR that follows expectations for a growing main sequence galaxy, before finally dropping of at around $z\approx0.5$.

\begin{figure}
  \begin{center}
    \includegraphics[width=.4\textwidth]{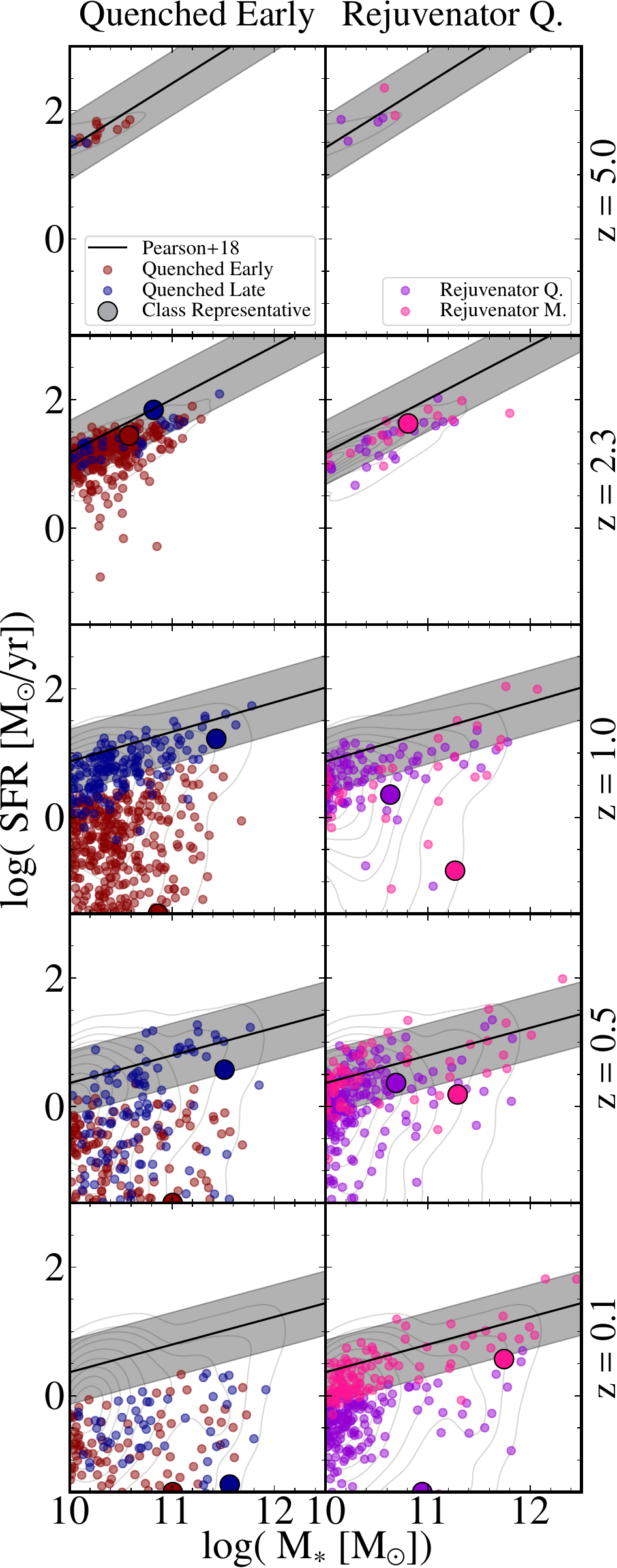}
    \caption{
        Similar to \autoref{fig:mst_p1e10} but focused on the difference between the subclass pairs of early- and late-quenched galaxies as well as quenched and main-sequence rejuvenators. 
        In this version, the colors refer to the subclass types instead of quenching incidents.
        The large points show the cherry picked representative galaxies from \autoref{fig:subclass_cherries}, again set to the bottom of the panels as half-circles for visibility in case of low SFR values.
        By construction, the difference between early- and late-quenched is most pronounced at $z=1$.
        At $z\approx 0$, the non-zero distributions are barely noticeable.
        Analogously, the difference between the two rejuvenator subclasses is the sharpest at $z\approx 0$.
        Both subclasses undergo rejuvenation cycles at different times, but the relation to the SFMS determines the classification.
    }
  {\label{fig:mssubclasses_1e10}}
\end{center}
\end{figure}

More and less rapid quenching scenarios are found among galaxies of both subclasses, as well as small and short-lived rekindling of star formation. To illustrate this distinction further, we plot in the left panel of \autoref{fig:mssubclasses_1e10} the star formation rate against stellar mass of galaxies from the "quenched" class at five different redshifts as in \autoref{fig:mst_p1e10}, but change the color scheme. Galaxies classified as early quenched are always plotted in red, while late quenchers are always in blue. This allows us to see that the majority of early quenchers still lie on or near the main sequence at $z=2.3$, such that their quenching must occur abruptly between $2>z>1$. Conversely, late quenchers do not surpass the mass range of the figure at $z=2.3$ and moment of quenching is less temporally confined, with occurrences spreading evenly between the time steps $1>z>0.5$ and $0.5>z>0$. This hints at different causes of quenching between early and late quenching, such as feedback and outflows as opposed to starvation or merger accretion. 

The median values of net gas accretion rates relative to dark matter assembly in \autoref{fig:classes_accrates} in the appendix support the image of an early starburst with a rapid transition toward net outflows for early-quenched galaxies. The relation between black hole mass and stellar mass shown in \autoref{fig:mbh_vs_mstar} in the appendix implies a correlation with AGN feedback, as early-quenched galaxies have the most massive black holes. As shown by \citet{kimmig:2025a}, an early and rapid formation also comes with an efficient gas transport towards the AGN. In addition to the starburst the feedback by the massive black hole then adds to the complete removal of gas leading to a permanent quenching. Late-quenched galaxies have lower accretion rate peaks and black hole masses, indicating a less rapid evolution due to gas accretion. The prevalence and impact of mergers in the two quenching scenarios will be disentangled in future work.

\subsection{Rejuvenating galaxies}\label{sec:rejuvenated}

Approximately $32\%$ of the galaxies in our sample are rejuvenators, and thus undergo one or more cycles of quenching and rejuvenating back onto the main sequence.
As highlighted by the changing colors in the third column of \autoref{fig:mst_p1e10}, this cycle can repeat up to three times, represented by a red-orange-yellow color scheme for the number of quenching transitions. We find that rejuvenating galaxies rarely quench completely.

The large range of different scenarios producing rejuvenation makes it difficult to find collective similarities in time-sensitive statistics. However, as we will discuss in \autoref{sec:imprints}, this class still reveals collective behaviour suggesting fundamental differences between formation conditions compared to the other classes.

\begin{figure*}
  \begin{center}
    \includegraphics[width=1.\textwidth]{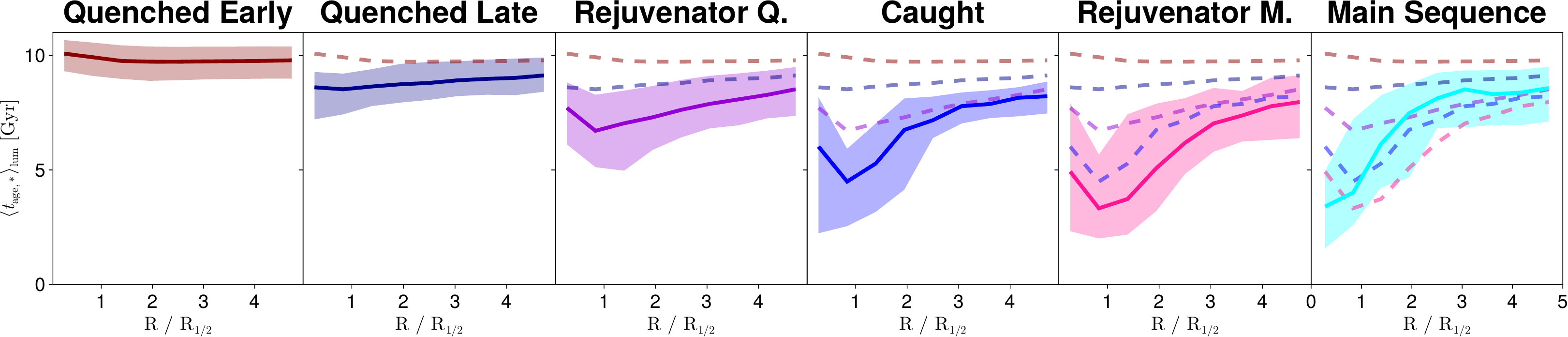}
    \caption{
        Median radial profiles and 68th-percentile spread of stellar ages at $z=0$.
        Before extracting the running median and scatter, we calculated the mean stellar ages for each galaxy in nine equal-width radial bins between $0$ and $5 R_\txt{1/2}$.
        The solid lines and shaded regions correspond to the statistics of the class indicated at the top of each panel.
        The dashed lines are the the running medians from the panels to the left of each panel for a direct comparison.
        Higher values correspond to an older population, which is fairly constant at around ten gigayears for early-quenched galaxies.
        Main-sequence-type galaxies show increasingly young populations towards the center, while the three types of rejuvenating classes are the youngest at roughly one half-mass radius and increase in age again towards the very center as well as the outskirts.
    }
    {\label{fig:radial_ages}}
    \end{center}
\end{figure*}
To further study the composition of active galaxies at $z \approx 0$, we differentiate rejuvenating galaxies that end up on the main sequence (in this work referred to as “Rejuvenator Main Sequence” or “RM”) from those that end up quenched (“Rejuvenator Quenched” or “RQ”). 
The conceptual difference is visualized with the chosen examples in the lower panel of \autoref{fig:subclass_cherries}, where we plot the sSFR against cosmic time of two example rejuvenator galaxies, one which ends up quenched (RQ, solid lilac line) and one which ends up star forming (RM, dashed pink line). 
Even though the quenched rejuvenator has a more consistent SFH, it drops off into a new quenched phase towards $z \approx 0$. 
The main-sequence rejuvenator undergoes an extended period of quiescence until rejuvenating back onto the main sequence in two rejuvenation cycles, ending up as actively star forming. 
In the right-hand column of \autoref{fig:mssubclasses_1e10}, we can see that the two subclasses of rejuvenators are not distinguishable by their location in the star formation rate versus stellar mass plot at higher redshifts, instead lying alternatively on or off the main sequence. 
We therefore conclude that there is no physical difference between the two classes of rejuvenators (unlike for the early and late quenchers), with the only distinguishing feature being the timing of their rejuvenation.
Intriguingly, we find that rejuvenating galaxies host less massive black holes (see \autoref{fig:mbh_vs_mstar} in the appendix), which may facilitate rejuvenation with less feedback emitted. We note also that the impact of mergers will be studied in future work using merger trees, which connect \subfind-identified progenitors of a galaxy through the simulation snapshots. This will allow us to look for differences between galaxy evolution classes regarding the time correlation between merger events and quenching or rejuvenation.

\subsection{Caught galaxies}\label{sec:caught}

A rare but nonetheless peculiar class of galaxies, namely $\sim 2\%$ of our sample, consists of such whose star formation rates drop below the main sequence at some point but never fully cease their star formation. 
Instead, they very continuously form stars with a nearly constant rate just below the main sequence, until they reappear on it once the overall main sequence itself has lowered to reach the SFR of these galaxies. An example case of this formation scenario is shown in the bottom panel of \autoref{fig:cherrytrack}.
We distinguish these galaxies from the rejuvenators, as they never truly rise in their SFR (i.e., rejuvenate), and instead are a result of the overall decline in cosmic SFRD.
Members of this galaxy class maintain a constant level of lower star formation rate, such that caught-type galaxies are especially promising candidates to test the equilibrium prediction of the “Bathtub” model \citep[][]{burkert:2017}. 
Therein, a constant gas accretion rate leads to an equilibrium state between accretion, star formation and feedback, resulting in a constant gas mass and star formation rate.
This formation scenario is particularly interesting since they are technically more stably retaining their star formation compared to the average main sequence galaxy, which is actually lowering its SFR with redshifts.
Their small sample size is expected as they continuously contribute to the scatter of the main sequence opposing the definition of a well-constrained star formation main sequence.

\section{Imprints of star formation histories}\label{sec:imprints}

A galaxy's star formation history depends on their evolutionary history, including both internal and external processes and leaving observable traces in their population of stars at $z=0$. Consequently, one could expect to find systematic differences in the galaxy properties of our different evolutionary classes. 

To test this, we plot in \autoref{fig:radial_ages} the median radial profile of the stellar age for all galaxies of a given class, normalized by their stellar half-mass radius. Indeed, we find noticeable differences among our galaxy classes.
Quenched galaxies (in the two leftmost panels) display very flat radial age profiles with $t_\txt{age} \approx 10\ \txt{Gyr}$, with only a faint trend for late-quenching galaxies towards younger stars in the center.
Galaxies that fully evolve along the main sequence have consistently younger stellar populations towards their centers, with the youngest average stellar population in the very center (rightmost panel).
Rejuvenating galaxies, especially those that are still on the main sequence at $z \approx 0$, produce a “u”-shaped stellar age profile, suggesting that rejuvenation tends to happen outside the stellar half-mass radii, similar to the results reported for high-redshift rejuvenation by \citet{remus:2023}.
For lower redshifts, \citet{hao:2024} find in MaNGA data a class of blue spirals with signs of rejuvenation, such as a dip in stellar age going from $0.5$~to~$1.5\cdot R_{1/2}$ (their profile type three), which is comparable to what we find here.
We note that we find a trend of continuously increasing stellar age with radius in the center only for main-sequence-type galaxies.
Caught-type galaxies show a similar age distribution as rejuvenating galaxies.
This could indicate a degeneracy of radial age profiles with respect to how sudden and steep the transition from below the main sequence to star-forming is.

Comparing the radial profiles of the different classes suggests an order of transitions, with early star formation starting of with flat profiles, and depending on how strong star formation is retained down to $z=0$ the central portion dips down toward younger ages. Quenched early and late have the least such prolonged star formation, with then RQ, then RM and caught are comparable, and finally those galaxies which are always on the main sequence show the strongest radial age gradient. 

\begin{figure}
  \begin{center}
    \includegraphics[width=.45\textwidth]{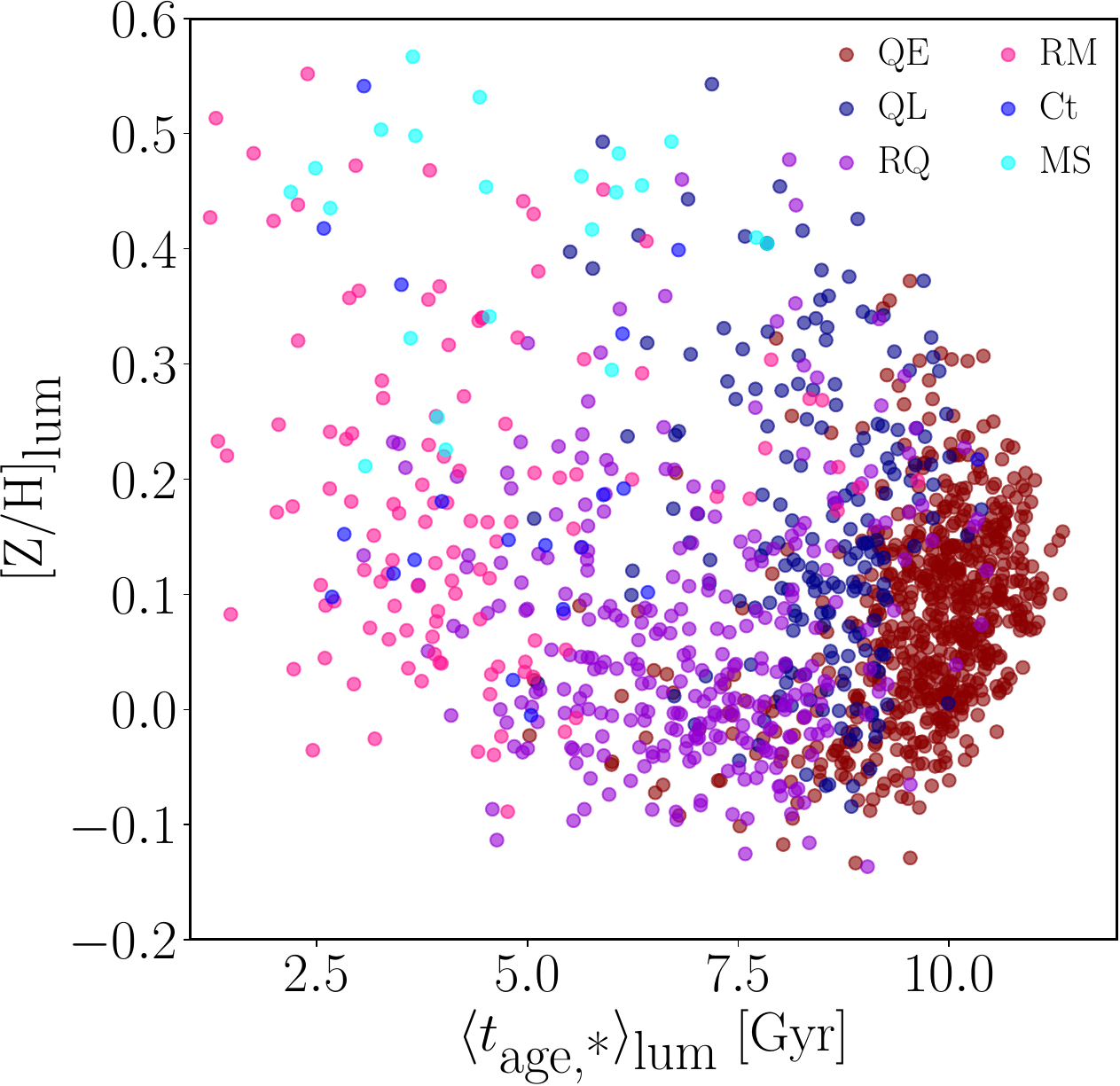}
    \caption{
        Mean metallicities and ages of the full galaxy sample at $z=0$.
        Less populated classes are in front of larger ones for visibility, while slightly decreasing the perceived scatter of the larger classes, like the early-quenched galaxy class. 
        This class concentrates at high ages and lower metallicities opposing main-sequence types.
        The intermediate classes of late-quenchers and rejuvenating galaxies bridge the gap in a dichotomous way of age and metallicity combination.
    }
  {\label{fig:met_ages}}
\end{center}
\end{figure}

Such an order of transitions between the classes also emerges from positions in the age-metallicity diagram of our galaxy sample as we show in \autoref{fig:met_ages}.
For all galaxies, their mean stellar metallicities $ \langle \txt{Z/H} \rangle $ are shown against their mean stellar ages $ \langle \txt{t}_{\txt{age},*} \rangle $.
We find that main-sequence-evolved galaxies represent the extreme end of high metallicities and young ages, while early-quenched galaxies are on the opposite end, strongly concentrated in the low-metallicity-old-age regime, with a significant scatter with respect to metallicity.

There emerges then an interesting difference between the late-quenched and rejuvenated-quenched galaxies, with both lying at similar mean stellar ages, but the former having much higher stellar metallicities than the latter. This implies that the stars formed in RQ galaxies were made of significant amounts of pristine gas, drawn from the cosmic web with little chemical enrichment, while late-quenchers continually enriched their gas and thus their newly formed stars, much like main sequence galaxies, before finally quenching. This is also why they lie in the transition between early-quenched and main sequence galaxies.

Curiously, though the RM-type galaxies are the overall youngest, together with the main-sequence galaxies, they tend toward much lower metallicities, comparable to their older RQ counterparts. It seems thus that low stellar metallicity in massive galaxies is found for two cases: either the galaxy formed and quenched rapidly in the early universe, or its star formation was not continuous, instead involving cycles of gas ejection, quiescence and re-accretion of more pristine, metal-poor gas that thus reduced the overall stellar metallicity. Conversely, remaining constantly on the main sequence requires some degree of gas retention, and thus leads to the overall highest stellar metallicities.

\begin{figure}
  \begin{center}
    \includegraphics[width=.46\textwidth]{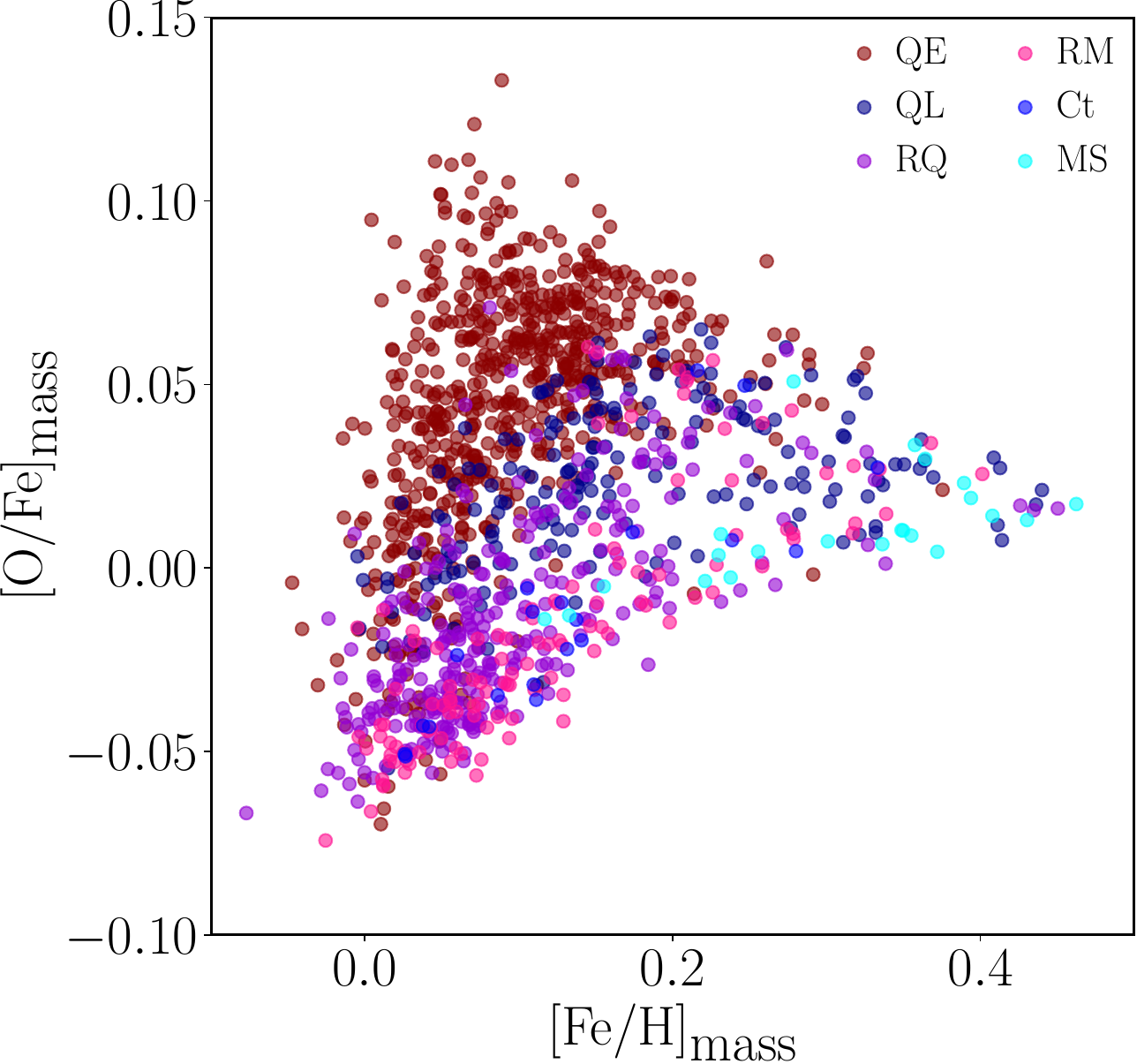}
    \caption{
    Mean stellar abundance ratios of the galaxy sample at $ z \approx 0 $. 
    Oxygen is chosen to represent $\alpha$ enrichment.
    Akin to \autoref{fig:met_ages}, different regions of the diagram are predominantly occupied by specific classes. 
    A clear gradient emerges from the top left to the bottom right of the distribution as QE-QL-RQ-Ct-RM-MS. 
    Main-sequence-type galaxies tend to the right-most corner of the distribution.
    }
    {\label{fig:OFe_FeH}}
    \end{center}
\end{figure}

The turbulent star formation histories with rejuvenation cycles also leave their imprint in the galaxies' stellar abundance ratios of oxygen to iron [O/Fe] and iron to hydrogen [Fe/H], which are measures of the fast $\alpha$ enrichment via type II supernovae and the slow iron enrichment via type Ia supernovae, respectively.
To demonstrate this, \autoref{fig:OFe_FeH} shows the mean abundance ratio relation [O/Fe] vs [Fe/H] of each galaxy.

We find that early-quenched galaxies (dark red) populate the highest [O/Fe] values.
Then the distribution transitions towards lower [O/Fe] and higher [Fe/H] values via late-quenched (dark blue) and then quenched rejuvenators (lilac). 
Main-sequence rejuvenators (pink), main-sequence evolved (cyan) and caught-type (blue) classes scatter around the bottom-right edge of the distribution, with rejuvenators at the lowest [O/Fe] values and main-sequence-types with the highest iron values [Fe/H], in agreement with \autoref{fig:met_ages} where we find them to boast the highest overall stellar metallicities.

Kimmig et al. (in prep.) show that the chemical evolution of galaxies in the \magpath simulations consistently begins at high [O/Fe] and low [Fe/H] values beyond the range of \autoref{fig:OFe_FeH}, with the initial enrichment dominated by type II supernovae and thus $\alpha$ elements.
The stellar metallicities of early-quenched galaxies are frozen in after a period of high star formation, leading to a final state with the highest [O/Fe] values since no additional stars with higher iron enrichment are formed.
The slower but steady contribution from type Ia supernovae for galaxies that have more time to evolve drives the distribution towards the lower values of [O/Fe] and higher values of [Fe/H] simultaneously. 
Interestingly, rejuvenating galaxies display the lowest $\alpha$ enrichment, which supports the picture of pristine gas accretion as the driver of rejuvenation.
The detailed treatment considering mass-dependent lifetime functions and metallicity-dependent stellar yields of AGB stars, SNII and SNIa, produce metallicities consistent with observations \citep{dolag:2017,kudritzki:2021,dolag:2025}. 
Since the dust and gas phases do not segregate on the galactic scale, we expect uncertainties in the IMF and stellar yields models (including for example the binary fraction) to dominate over the impact of dust.
Nonetheless, the values shown in this work are to be compared to dust-corrected results.

\begin{figure}
  \begin{center}
    \includegraphics[width=.45\textwidth]{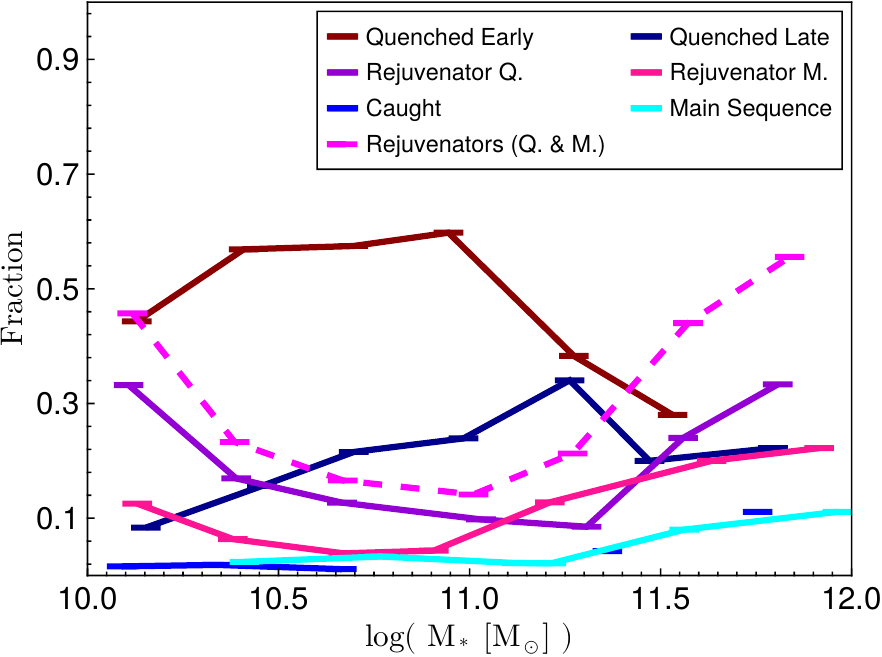}
    \caption{ 
        Fractions of formation class members in stellar mass bins at $z\approx 0$. 
        Slight shifts of data points originate from taking the median stellar masses instead of bin centers. 
        Early-quenched galaxies (dark red) peak at around $M_* = 10^{11}$, late-quenched ones (dark blue) at slightly higher masses. 
        Galaxies with rejuvenation cycles (dashed magenta line) in their formation histories are more prevalent on the lower and higher mass ends of the range between $10^{10} M_\odot \leq M_* \leq 10^{12} M_\odot$.
        Quenched rejuvenators (lilac) tend more towards lower masses, while a rejuvenation sustained until $z\approx 0 $ (pink) is more often found at higher masses.
        Main-sequence (cyan) and caught types (blue) are generally rare with a faint preference for the high-mass end.
        Both subtypes are more prevalent on the lower and higher mass ends of the range between $10^{10} M_\odot \leq M_* \leq 10^{12} M_\odot$. 
    }
  {\label{fig:mstars_relative}}
\end{center}
\end{figure}

We show in \autoref{fig:mstars_relative} that producing stars in rejuvenation cycles not only affects the observable combination of age and metallicity, but also correlates with the stellar mass of the galaxy at $z=0$. We plot there the fractions of galaxies that belong to each class within a given stellar mass bin.

Interestingly, we find that rejuvenating galaxies (dashed magenta line) lie predominantly at either the low- or high-mass end, which is in part because the overall largest group, the early-quenching galaxies (dark red), concentrates around $10^{10.5}-10^{11}\ \txt{M}_\odot$. By contrast, late-quenching galaxies (dark blue) tend toward slightly higher stellar masses. However, we note that the “u”-shaped rejuvenator distribution could imply two separate physical mechanisms which dominate. On the right, the increasing rejuvenator fraction with increasing stellar mass may be the result of an increasing number of mergers within their evolutionary history (as the most massive objects have experienced the most mergers), as well as the possibility of cooling from the hot halo \citep[e.g.][]{fabian:1977:nulsen, sarazin:1986,gonzalez_villalba:2025}.
To investigate the cause of the bimodal distribution, splitting the rejuvenators up by their stellar mass into high- and low-mass subclasses should reveal collective differences in gas accretion evolution, in-situ fractions and merger histories. The latter would be a plausible main driver for rejuvenation of the higher-mass galaxies for which smooth accretion is quenched by shock heating by their hot halos \citep[e.g.][]{dekel:2006:birnboim}. Furthermore, explicitly tracing gas particles would provide further insight into the different mass assembly histories, which is a topic for future work.

For the low mass end at $M_*\approx10^{10}M_\odot$, we find that quenched-early and quenched-late account for $52\%$, which increases up to $70-80\%$ by $M_*\approx10^{11.25}M_\odot$. This is because gas falling into halos with a virial mass greater than $M_\txt{vir} \gtrsim 10^{12} M_\odot$ (galaxy stellar masses of $M_* \gtrsim 3\times10^{10} M_\odot$) gets shock-heated to high temperatures where cooling becomes inefficient \citep[e.g.][]{rees:1977:ostriker}. This leads to the apparent knee in the galaxy luminosity function \citep[][]{schechter:1976} and stellar mass function \citep[e.g.][]{bullock:2017:boylan-kolchin}. Consequently, it is the increase in those groups which leads to the decrease in rejuvenated galaxies.

However, simply becoming quenched does not mean a galaxy may not rejuvenate. Instead, we note that the lower rate of rejuvenated galaxies toward intermediate mass galaxies may be required by construction, as to reach stellar masses of $M_*>10^{10.5}M_\odot$ requires a high SFR over a long period of time, with simply insufficient time to both quench and rejuvenate. Indeed, the late-quenched galaxies are those which rise the strongest with increasing stellar mass, meaning that they represent galaxies which grew significantly in stellar mass by forming it in-situ, before finally running out of gas (or quenching via a dry merger). Simultaneously, low mass galaxies are more easily quenched as they have more shallow potential wells \citep[e.g.][]{dekel:1986:silk,efstathiou:2000}, and may therefore be more susceptible to short-term ejection and re-accretion of gas, and galaxies which end at lower masses must have on average spent a longer time with such shallow potential wells compared to more massive galaxies.

Generally, we note that rejuvenated galaxies exist at all stellar masses, while our main-sequence galaxies (cyan), though few in number, show a clear trend toward higher $z=0$ masses. This is a direct consequence from the constraint of lying on or near the main sequence at all times. When taking an initial galaxy of $M_*=10^8M_\odot$ at $z=4,3,$ and $2$, assuming a Chabrier IMF \citep{chabrier:2003} with mass loss following \citet{leitner:2011} and determining its star formation rate in time steps of $10\,$Myr based on the evolving \citet{pearson:2018} main sequence, we determine the expected final $z=0$ mass as $M_*=6\times10^{10}M_\odot,\, 4\times10^{10}M_\odot$~and~$3\times10^{10}M_\odot$. 

We note that this is assuming an ex-situ fraction of $0\%$, such that the true mass is likely higher, and also that assuming a \citet{kroupa:2001} IMF would only increase the final mass, due to the higher fraction of low mass stars. Consequently, it is essentially by construction that observed lower mass galaxies at $z=0$ must have undergone some periods of quiescence, or must have formed incredibly late in cosmic time (after the peak of the cosmic SFRD).

Another alternative for such lower stellar mass galaxies is belonging to the caught class, which we only find toward the low-mass end of our distribution (blue line in \autoref{fig:mstars_relative}). By retaining a stable star formation rate just below the main sequence, these galaxies never strongly quench and rejuvenate but still end up with lower overall masses. 

Combining the results from \autoref{fig:radial_ages}, \ref{fig:met_ages}, \ref{fig:mstars_relative} and the last row of \autoref{fig:mst_p1e10}, we find that the different star-formation histories leave distinct imprints on galactic features at redshift $z=0$. 
As a consequence, we can derive a likely star formation history of a galaxy using one or more of these features, which we will study in more detail in a forthcoming paper.

\section{Gas flows fueling star formation}\label{sec:accretionrates}

We then turn to understand the cause of these different evolutionary histories of galaxies. 
Stars form from giant molecular clouds (GMCs) surpassing the Jeans criterion \citep[][]{jeans:1902}. 
While the composition of gas phases plays an important role for star formation, the rates fundamentally depend on the amount of gas available to a galaxy.
This reservoir is built up and replenished by gas accretion from the environment and depleted by star formation and outflows. 
Gas flows and star formation can be directly related through the continuity equation assuming conservation of mass, as defined for example by \citet{bouche:2010} in their equation (3):
\begin{equation}\label{eq:galconti}
    \begin{aligned}
        \Dot{M}_\txt{molgas} & = -(1+\eta-R)\ \txt{SFR} + \Dot{M}_\txt{acc}
    \end{aligned}
\end{equation}
Here, $\eta$ quantifies the outflow mass loading by supernovae, and R describes the gas return through stellar winds, each relative to the star formation rate.
In case of constant accretion rates and outflows dominated by stellar feedback with constant feedback parameters the “Bathtub” model predicts an equilibrium state with a gas mass and star formation rate proportional to the accretion rate \citep[e.g.][]{burkert:2017} when star formation and feedback cancel out gas accretion.
This final galaxy gas reservoir determined by a constant accretion rate serves as the volume of the figurative bathtub.
Such an equilibrium state is expected to settle in roughly within the depletion time $t_\txt{depl} \equiv t_\txt{ff} / \epsilon_\txt{ff}$. Theoretical models have generally found $\epsilon_\txt{ff} = 0.01$ \citep[][]{krumholz:2005,krumholz:2007}, which is broadly mirrored in observations \citep{lada:2010,heiderman:2010,krumholz:2012,utomo:2018}. For giant molecular clouds with $n_\mathrm{GMC}\approx30 \mathrm{cm}^{-3}$, assuming a mean molecular mass $\mu_\mathrm{GMC}=2.34\times10^{-24}$g \citep{krumholz:2012}, we calculate the free fall time as $t_\txt{ff,GMC}=\sqrt{3\cdot\pi/(32\cdot \txt{G}\cdot \rho_\mathrm{GMC})}\approx 8\,$Myr. Correspondingly, $t_\txt{dep,GMC}\approx 0.8\,$Gyr, which may be shorter for denser clouds.

On the other end, for full galaxies at $z=0$ the depletion timescale is longer at $t_\txt{dep,gal}\approx 2\,$Gyr \citep{utomo:2018}, with the IGM generally more diffuse than the denser GMCs. However, this changes across redshift, as high redshift galaxies are much more gas rich and therefore closer to the limit of GMCs. Thus, we choose as our smoothing window for star formation rates a value in-between of $t_\mathrm{smooth}=1\,$Gyr.

We measure in- and outflow rates of galaxies from particle data in the simulation and derive expected star formation rates according to \autoref{eq:galconti}.
The flow rates are derived by drawing a thin spherical shell around each galaxy at a radius of $r_\txt{gal} = 0.1\times r_\txt{vir}$ and a thickness of $\Delta r = 0.01\times r_\txt{vir}$.
Each SPH gas particle contributes to flow rates weighted by its mass, velocity and fraction of its volume overlapping with the shell volume around the respective galaxy.
Similar to the work by \citet{mitchell:2020}, we require particles to have a radial velocity of $|v_\txt{rad}| < 50\ \txt{km/s}$ to contribute to gas flow rates.
For this work, we do not differentiate between outflow by stellar or AGN feedback. 
Instead, we set $\eta = 0$ and use the net accretion rate $\Dot{M}_\txt{net} \equiv \Dot{M}_\txt{in}-\Dot{M}_\txt{out}$ for gas accretion. 
The return fraction is assumed constant as $\txt{R} = 0.4 $ \citep[][]{bruzual:2003:charlot}.
With this implementation we can derive the evolution of the gas reservoir and respective star formation rates purely from gas accretion rates.
Further, we solve the differential equation numerically in order to account for changes in accretion rates and derive gas masses as
\begin{equation}
    \begin{aligned}
    M_{i} & = M_{i-1} + \left( -(1+\eta-R)\ \frac{\epsilon_\txt{ff}}{t_\txt{ff}}\ M_{i-1} + \Dot{M}_\txt{acc} \right)\ \Delta t
    \end{aligned}
\end{equation}
with an expected star formation rate of $ \txt{SFR}_{i} = M_{i} / t_\txt{depl}$.
This numerical approximation requires a time resolution with $\Delta t \ll \frac{t_\txt{ff}}{\epsilon_\txt{ff}\ (1+\eta-R)}$.
Since the particle data output does not meet this requirement, we interpolated linearly between gas flow rates measured in available snap shots. 

With this method to compensate for the limited time resolution, we effectively assume that accretion rates do not fluctuate on shorter time scales than the particle data snapshot step size of $\sim 200-300$ Myr.
Fluctuations on smaller time scales should also not impact the derived SFR values, since they would be smoothed out by the time scale of the system determined by the “Bathtub” model's time scale, which we set at the lower limit of GMC depletion times $t_\txt{BT} = (1+\eta-R)\ \frac{\epsilon_\txt{ff}}{t_\txt{ff,GMC}} \approx t_\txt{depl} \approx 0.5\ \txt{Gyr}$.

We compare the resulting measured inflow rates to expectations from dark matter assembly and smooth accretion according to \citet{dekel:2009} by normalizing our measurements by their halo-mass-dependent and redshift-dependent results. 
We find that our measured absolute inflow rates are higher by a factor of roughly two to five at the virial radii, with net accretion rates (inflow minus outflow) lying at similar values to the model expectation, both at the virial radius as well as onto the galaxy itself -- see \autoref{fig:classes_accrates} in the Appendix, as well as Seidel et al. (in prep.). Deviations toward higher or lower net accretion are to be expected, as there must be additional gas mixing and ejection that go beyond simpler dark matter inflow assumptions. 

On the total net accretion onto the galaxy, we note here only that all classes begin with comparable net gas accretion rates at high redshifts, with the quenched-type galaxies peaking early at high net values and then fall down to zero. The other classes retain significant net inflow, that smoothly declines toward low redshifts. For a deeper discussion, see \autoref{fig:classes_accrates} in the Appendix.

Instead, here we ask how efficient the galaxies of each class are at funneling the total inflow of gas onto the halo (at $r_\txt{vir}$) into the galaxy itself (at $r_\txt{gal}\equiv0.1\cdot R_\txt{vir}$). Because we wish to capture if there is significant mixing within the halo itself occurring, we consider here the absolute inflow rates instead of the net accretion (as gas fountains and cycles can result in net zero accretion, even though there is significant gas mixing).

\begin{figure}
  \begin{center}
    \includegraphics[width=.45\textwidth]{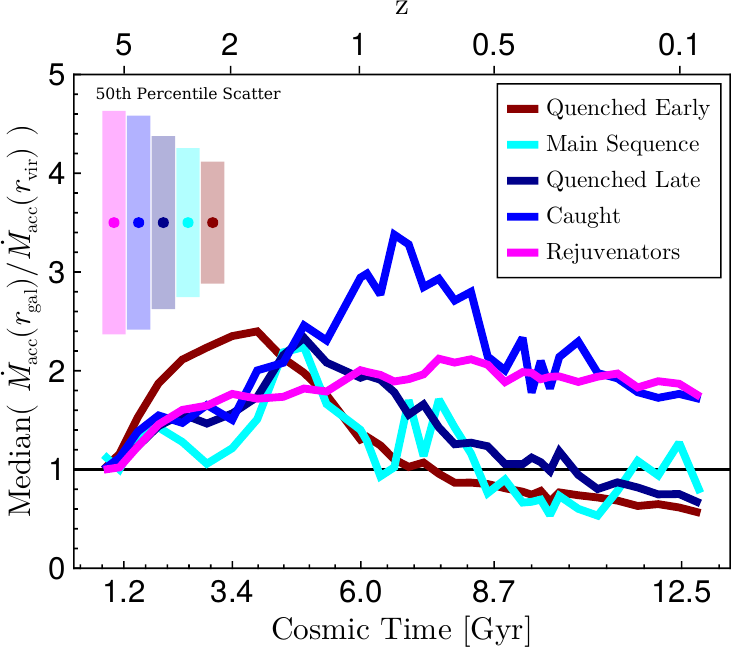}
    \caption{
    Evolution of median relative accretion rates of each class across cosmic time. 
    The underlying values were retrieved by calculating the ratio of accretion rates at $0.1 \cdot R_\txt{vir}$ and $R_\txt{vir}$ for each galaxy at each particle snapshot. 
    Early-quenched galaxies exceed at accretion rates onto the center at high redshifts, while abruptly decreasing after redshift $z<2$.
    Main-sequence and late-quenched types have a similar evolution but with a clearly delayed peak. 
    Rejuvenators do not peak but rather reach a plateau, while caught-types reach the same plateau after a strong delayed peak.
    \label{fig:accgal_by_accvir}}
    \end{center}
\end{figure}
We plot in \autoref{fig:accgal_by_accvir} the median ratios of gas inflow rates onto the halos and onto the galaxies against cosmic time for each class of galaxies (colored lines). Interestingly, we find that overall infall rates are generally higher in the centers of halos than at their boundary for all classes (with equality in the ratios shown by the solid black line). This implies that all galaxies cycle some amount of gas within their halo, which allows the same gas to be accreted (and ejected) multiple times onto the galaxy, leading to an inflow ratio greater than one. While this could also be due to the time delay of infalling gas at the virial and galactic radii, the consistently higher net accretion rates at $r_\txt{vir}$ (see \autoref{fig:classes_accrates}) for the two classes support the cycling hypothesis. Crucially, however, clear differences emerge in the trends over time for the different evolution classes.

Early-quenching galaxies exceed at funneling gas into the center in the early universe, while at later times not even all of inflowing gas onto the halo makes it into the galaxy. The high peak at early times (both in absolute flow rate, see \autoref{fig:classes_accrates}, as well as significant mixing in the halo itself shown in \autoref{fig:accgal_by_accvir}) may indicate a star burst at early times, which would both rapidly deplete as well as eject the gas, similar to what is shown by \citet{kimmig:2025a}.

With a delay the same behaviour is displayed by late-quenching galaxies and even main-sequence-type galaxies. This means that all these galaxies to not predominantly cycle the same gas within their halos (where the ratio in \autoref{fig:accgal_by_accvir} would then be greater than~1), but instead approximately transport all or nearly all of the inflowing gas into the center, where it also remains. The difference is that for the quenched galaxies, this is practically no gas at all, while for the main sequence galaxies there is a significant amount, but the amount of transport from halo boundary to center is comparable among all three classes. Main-sequence-type galaxies thus do not cycle, but simply accrete and retain, with long depletion times and little feedback to eject their gas again. 

By contrast, caught- and rejuvenator-type galaxies set themselves apart from the other classes with median values higher by a factor of two towards $z\approx 0$. This means that at all times they are cycling significant amounts of their gas within their own halo, potentially fueled by fountain processes that restore high accretion rates in the center while accretion at the halo outskirts declines. Cycling must on average then result in lower star formation rates over the entirety of cosmic time, as the main-sequence-type galaxies are generally more massive than both the caught and rejuvenated galaxies. 

\begin{figure}
  \begin{center}
    \includegraphics[width=.45\textwidth]{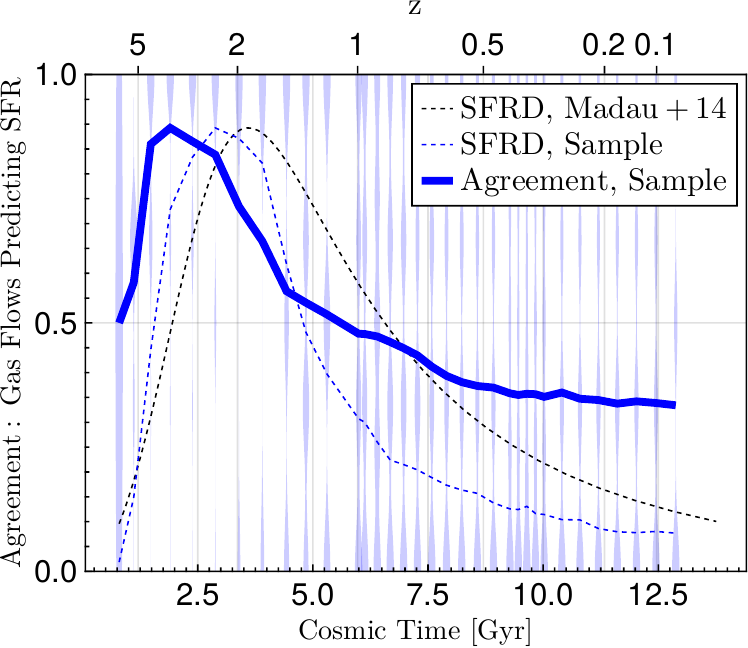}
    \caption{
        Violin plot indicating the mean agreement of expectations from net accretion and the star formation rates derived from \autoref{eq:galconti} with simulation output. 
        An agreement is registered when both values deviate less than 0.5 dex from each other or for quenched galaxies, when simulation output and derived values return any SFR below $10^{-3}\txt{M}_\odot / \txt{yr}$.
        The vertical histograms (violins) count the agreement fraction of each galaxy within $500\ \txt{Myr}$ before and after each snapshot, combining three to four snapshot data points.
        This produces discretized fractions accumulating at 0.0, 0.25, 0.3, 0.5, 0.7, 0.75 and 1.
        The solid line follows the mean values within each bin.
        The blue dashed line shows the evolution of the star formation rate density of the galaxy sample normalized to the range of the figure.
        The black dashed line shows observational results for the evolution of the star formation rate density compiled by \citet{madau:2014:dickinson} and was normalized as well.
    }
    {\label{fig:BTvsSUBFIND}}
    \end{center}
\end{figure}

In order to investigate how well the evolution of star formation rates is determined by gas accretion, we show in \autoref{fig:BTvsSUBFIND} the agreement between the expected star formation obtained from the “Bathtub” model using the measured gas accretion rates, and the real simulation output for the total galaxy sample across cosmic times. Note that we set the mass loading factor $\eta$ to zero, as we directly measure the outflow rates, accounting therefore for both the contributions from AGN feedback and stellar feedback.
The violins indicate the fractions of galaxies that have star formation rates in agreement with expected values when using \autoref{eq:galconti}.
We find that the fraction of “Bathtub”-like galaxies in the total galaxy sample (solid blue line) quickly rise and peak soon after $z=5$.
This means that at high redshifts, the assumed constant values for the return fraction $R$ and depletion time $t_\txt{depl}$ of a dense giant molecular cloud in a galaxy with a star formation efficiency of $\epsilon_\txt{ff} = 0.01$ describe the transformation of accreted gas into stars well. Going toward lower redshift, then, the agreement drops. We note that there are more elaborate descriptions of the depletion time scale, like a redshift-dependent description following the trend and offset to the main sequence $\Delta \txt{MS}$ shown by \citet{tacconi:2020}, yet it is noteworthy that our simple approach works well at high redshift.

The shape of this agreement is reminiscent of another well known relation. 
Thus, for comparison, we plot in arbitrary units (as we care about the shape and peak) the star formation rate density (SFRD) of the sample (dashed blue line), which peaks just before $z=2$, in good agreement from the moment of SFRD peak according to observations \citep[][]{madau:2014:dickinson}. We find a similar form, and slightly earlier peak, which may indicate that one of the reasons for the observed cosmic SFRD is the efficiency at which present gas can be converted into stars, with high redshift galaxies following the “Bathtub” model of direct conversion on short depletion times, while lower redshift galaxies grow increasingly inefficient.
This is consistent with previous findings by \citet{teklu:2023} where the overall growth in size leads to lower star formation rates with the same amount of cold gas available. In future work, we plan to more directly explore the link between the internal gas distribution of galaxies, a detailed characterization of their accretion flows (e.g. smoothness and angular momentum), and the resultant agreement with the “Bathtub” model.

\section{Environmental impact}\label{sec:environment}

It is well established that environment plays an important role for the star formation of satellites \citep{bahe:2017,lotz:2019}, but as we consider here only galaxies which remain as centrals until $z\approx0$ the picture is less clear. In this case, \citet{remus:2023} and \citet{kimmig:2025a} do find indications that the environment plays a crucial role both for quenching as well as for rejuvenation, at least at high redshifts.
This can be understood as star formation histories being dependent on both the availability of surrounding gas and the interaction with neighboring galaxies.
A large number of lower-mass neighbors could fuel star-formation through minor mergers, while higher-mass neighbors could steal away ejected or loosely-bound gas and quench star formation in a similar fashion as for the Roche lobe of binary stars \citep[e.g.][]{morris:1994}.

Filaments have been proven as effective means to transport gas into the center while shielding against the quenching effect of a shock-heating hot halo \citep[][]{dekel:2009}.
While quenching can be induced by the intrinsic properties of a galaxy via hot halo and AGN, the frequent occurrence of rejuvenation processes suggests that the star formation history is at least in part determined by the environment.

\citet{seidel:2025} have shown that the \magpath simulations are well-suited to study gas flows between halos and their environments, finding agreement in the flow-rates across both box resolution and size. While the gas mass resolution here is not high enough to trace the evolution of individual molecular clouds, we well resolve the more massive gas flows which are needed to rejuvenate galaxies in the mass ranges considered in this work. We therefore search for systematic differences between the environments of the classes with respect to the following two parameters:
\begin{itemize}
    \item The $d_5$ parameter quantifies the density of the environment, defined as the minimum radius at which five neighboring galaxies, each with a total mass greater than $10^{11}\ M_\odot$, are found \citep[][]{remus:2023}.
    \item The asphericity of gas infall at the virial radius. This parameter is derived by first performing a multipole expansion on the accretion rate distribution on a spherical shell and then calculating a ratio of the power of multipoles divided by the monopole power (Seidel et al. in prep). Lower values (i.e., high monopole powers) correspond to isotropic accretion.
\end{itemize}

For the former, we find only mild differences, and as such refer the interested reader to \autoref{fig:violin_d5} in the appendix for a further discussion. Here, we only briefly mention that caught galaxies tend toward slightly under-dense environments, while main-sequence-type show the opposite trend, with rejuvenated and quenched galaxies in the middle.
However, these deviations are small, suggesting that our classes of galaxies are independent of the galaxy density of its environment.

\begin{figure}
  \begin{center}
    \includegraphics[width=0.48\textwidth]{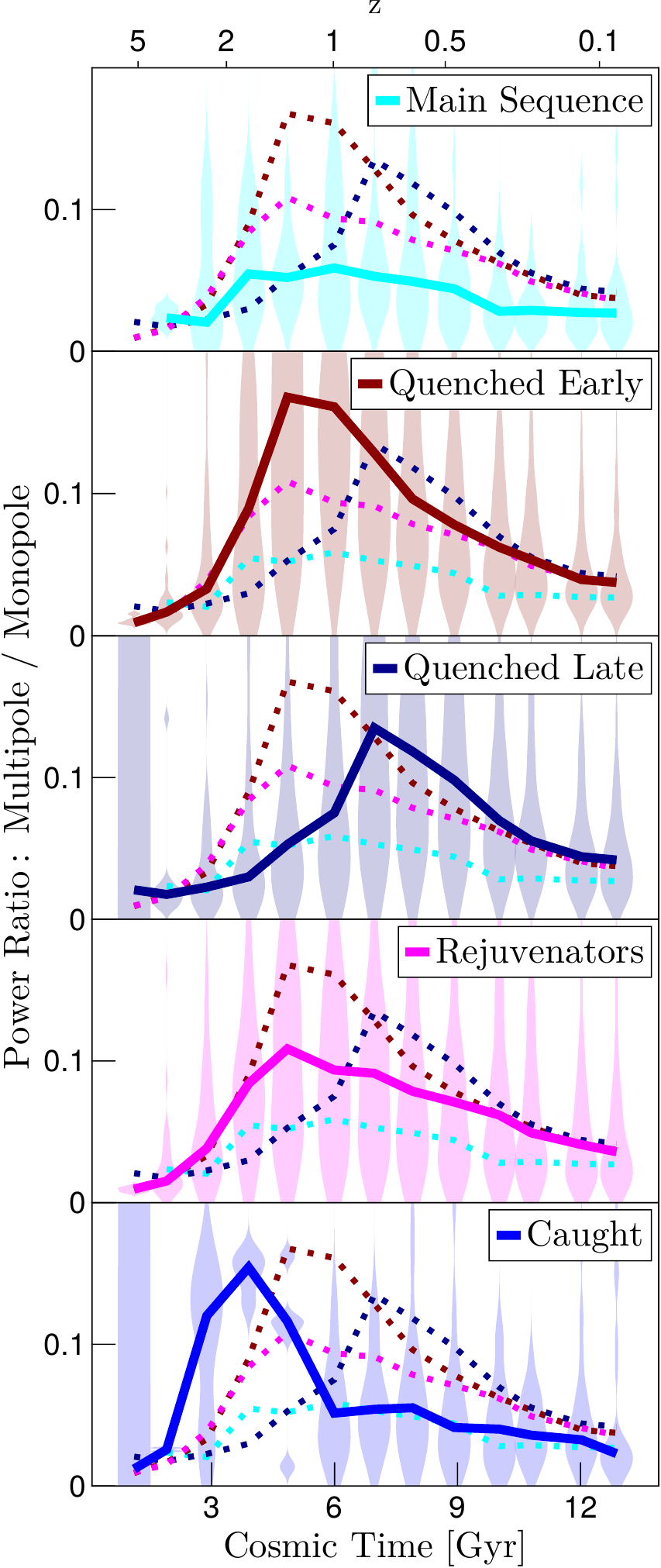}
\caption{
    Violin distributions of the evolution of asphericities for each class at each snapshot, with main sequence rejuvenators and quenched rejuvenators combined as “Rejuvenators”.
    The asphericity of gas infall is evaluated on the spherical surface of each virial radius.
    The solid lines follow the median of each snapshot. 
    The dotted lines are the median distributions of the other classes for a direct comparison. 
    The main-sequence types maintain lower values and therefore a predominantly spherical gas infall.
    The other classes display peaking distributions between redshifts $2 > z > 0.5$.
	}
	\label{fig:asphericities}
    \end{center}
\end{figure}

Instead of this total density, it may be the geometry of the gas inflow. Indeed, we find a more pronounced difference between the classes when tracking how gas infall asphericities at their virial radii evolve.
\autoref{fig:asphericities} shows the evolution of median asphericities for each formation class as well as the underlying distributions as violins.

We find that galaxies which stay on the main sequence (cyan) also experience consistently isotropic infall throughout cosmic time, with only slightly higher asphericity between $2 > z > 0.5$.
The other classes undergo a pronounced peak of aspherical infall, each at a different redshift.
Caught-type galaxies are the first class to peak in asphericity at $z \approx 2$.
Next are early-quenched galaxies which peak at $z \approx 1.2$.
Last to peak in asphericity as a group are the late-quenched galaxies at $z \approx 0.7$.
Rejuvenating galaxies peak at the same time as early-quenched galaxies, but with much lower total asphericity, indicating an infall history more similar to main-sequence-type galaxies.

This peaking distribution suggests a typical formation history of overall first isotropic infall, then a period of more filamentary accretion followed by a more isotropic flow that could reflect the buildup of halo fountains. Merger rates decline toward lower redshifts \citep{oleary:2021}, such that increasing isotropy must be the result of smooth accretion. For the case of the main-sequence type galaxies, the rates are high enough (see \autoref{fig:classes_accrates}) that this gas is likely the result of increasing amounts of ejected gas inflowing back from all directions. 
We will study this process of fountain build-up in more detail in a future study.

\section{Summary and conclusions}\label{sec:summary}

We trace the evolution of $ 1282 $~central galaxies in the highest-resolution box of the \magpath cosmological hydrodynamical simulation suite through cosmic time, especially focusing on their gas inflow and star formation rates. 
By comparing their star formation rates with respect to expectations from star forming main-sequence observations by \citet{pearson:2018}, we identify moments of sustained quenching and rejuvenating and classify the formation histories accordingly.

We find four characteristic formation pathways with respect to the SFR over cosmic time for central galaxies with $M_* > 10^{10}\ \txt{M}_\odot$ at $z\approx 0$:
\begin{itemize}
    \item Around $50\%$~(646) quench soon after the peak of star formation at $z\approx 2$, while $15 \%$~(187) quench later between $1>z>0$.
    \item Only $2\%$~(22) evolve on the star formation main sequence until $z\approx 0$, where they make up $14\%$ of galaxies on the scaling relation. The other $86\%$ has undergone quiescent periods.
    \item Another $2\%$~(20) of galaxies retain a lower but constant star formation rate, and reappear on the main sequence as it declines towards lower redshifts without need for rejuvenation.
    \item A large fraction of $32\%$~(407) of galaxies undergo one or more cycles of rejuvenation on gigayear timescales, with $9 \%$~(117) of all galaxies returning onto the main sequence by $z\approx 0$, while the other $23\%$~(290) rejuvenate at some point but end up as quenched by $z\approx0$.
\end{itemize}
This means that the dominant fraction of galaxies ($117$~of~$159$) found on the star formation main sequence at $z \approx 0$ have undergone cycles of quenching and rejuvenation.
These classes exist independent of which main sequence definition is used, for all main sequence definitions studied in this work \citep[][]{speagle:2014, pearson:2018, leslie:2020, popesso:2023}, and only the relative frequencies per class change with definition.

These formation scenarios leave characteristic imprints on mean stellar ages and metallicities as well as the radial distribution of stellar ages, with rejuvenation generally taking place beyond the stellar half-mass radius.
Galaxies that evolve along the main sequence consistently also have the most efficient star formation in their very centers, and the highest stellar metallicities.
The final black hole masses at $z \approx 0$ of main-sequence evolved galaxies and rejuvenating galaxies are distinctly lower than early-quenched galaxies at any given stellar mass, especially on the lower-mass end of the sample. Early formation conditions of efficient gas transport to the galaxies' centers enhances not only star formation but also black hole growth, which in turn may impede rejuvenation or an evolution along the main sequence due to enhanced AGN feedback.

We also find that the scenarios emerge independent of the densities of galactic environments when looking at the $d_5$ parameter, with no significant differences found between the formation classes across all redshifts.
However, a clear distinction is possible when looking at the evolution of the asphericity of the infalling gas at the virial radii. Early and late quenching galaxies strongly peak in anisotropic accretion around the same time when their star formation rates drop, in stark contrast to the absence of such an asphericity peak for main-sequence evolving galaxies. 
Explaining the physical reasons for this correlation is subject to future studies.

We compared the simulated star formation rates to predictions assuming mass conservation and a constant depletion time. The agreement between simulation results and expectations from gas flows evolves similar in shape and time to the evolution of the cosmic star formation rate density, with early redshifts showing the best agreement that then declines toward lower redshifts. Consequently, at $z>2$, the conditions of the Universe are described well by a simple description of star formation derived from gas flows and a depletion time of $t_\txt{depl}\approx 0.5$ Gyr. 
After that time, the conditions change, as only about $40\%$ of galaxies' star formation rates are consistent with expectations from the “Bathtub” model equation, which may be due to a more complex phase where star formation is less dependent on the net gas accretion, and additional physical processes (heating, turbulence, dry mergers) become more dominant in regulating star formation.

We conclude that galaxies evolving consistently along the main sequence are relatively rare, and most current day main sequence galaxies have undergone one or multiple cycles of quiescence and rejuvenation.

\begin{acknowledgements}
We thank the referee for their valuable feedback which helped with clarification and context of the results.
This work was performed at the LMU Munich with the support by the Excellence Cluster ORIGINS.
The \magneticum simulations were performed at the Leibniz-Rechenzentrum with CPU time assigned to the Project {\it pr83li}. This work was supported by the Deutsche Forschungsgemeinschaft (DFG, German Research Foundation) under Germany's Excellence Strategy - EXC 2094 – 390783311. We are especially grateful for the support by M. Petkova through the Computational Center for Particle and Astrophysics (C2PAP). LCK acknowledges support by the DFG project nr. 516355818. KD acknowledges support by the COMPLEX project from the European Research Council (ERC) under the European Union’s Horizon 2020 research and innovation program grant agreement ERC-2019-AdG 882679.
\end{acknowledgements}

\bibliographystyle{aa}
\bibliography{biblio}

\appendix

\section{Varying the smoothing window}\label{sec:A_variationbywindow}

\FPeval{\resultpms}{round(34/212,2)}
\FPeval{\resultpc}{round(6/20,2)}
\FPeval{\resultpr}{round(46/123,2)}
\FPeval{\resultpq}{round(256/630,2)}
\FPeval{\resultpms}{round(34/212,2)}
\FPeval{\resultpms}{round(34/212,2)}
\FPeval{\resultpms}{round(34/212,2)}
\FPeval{\resultpms}{round(34/212,2)}
\FPeval{\resultpms}{round(34/212,2)}
\FPeval{\resultpms}{round(34/212,2)}
\FPeval{\resultpms}{round(34/212,2)}
\begin{table}[h!]
\begin{centering}
    \caption{Numbers of galaxies within each formation class with respect to expectations according to the SFMS as found by \citet{pearson:2018}.
    }
\begin{tabular}{|*{4}{c|}}
    \hline
     & 0.5 Gyr & 1 Gyr & 1.5 Gyr \\ \hline
    MS.& \gradient{13} & \gradient{22} & \gradient{32} \\
    Q. & \gradient{645} & \gradient{833} & \gradient{907} \\
    R. & \gradient{615} & \gradient{407} & \gradient{297} \\
    C. & \gradient{20} & \gradient{20} & \gradient{20} \\ \hline
\end{tabular}
    \tablefoot{
        From top to bottom: main sequence, quenched, rejuvenating and caught. We compared the results using three smoothing windows of $0.5$, $1.0$ and $1.5$ gigayears each are compared in order to test the robustness of the classification.
    }
  {\label{tab:classify_smoothing}}
  \end{centering}
\end{table}

We classify evolution scenarios as shown by representative examples in \autoref{fig:cherrytrack} algorithmically. 
In order to tie the star formation rates to gas flows we applied a smoothing window of one gigayear to the simulation output representing the depletion time scale. 
The depletion time scale roughly determines how fast a galaxy globally reacts to change in gas flows according to the “Bathtub” model (\autoref{eq:galconti}).

In order to check for robustness, we applied the algorithm using also a 0.5 and a 1.5 Gyr time window.
The effect on sample sizes is shown in \autoref{tab:classify_smoothing}.
Larger smoothing windows disregard short-term fluctuations of star formation rates. 
As a consequence, more galaxies are considered main-sequence or quenched. 
Those would otherwise have been attributed to the group of rejuvenating galaxies, as short-term fluctuations would have been considered quenching and rejuvenating processes. 
The number of caught-type galaxies remains stable, as they have been classified manually. 
However, despite the varying class member numbers, each class roughly maintains its size, showing these characteristic evolution pathways emerge for a range of smoothing window size choices.

\section{Gas flows}\label{sec:A_gasflows}

\begin{figure}
  \begin{center}
    \includegraphics[width=.9\columnwidth]{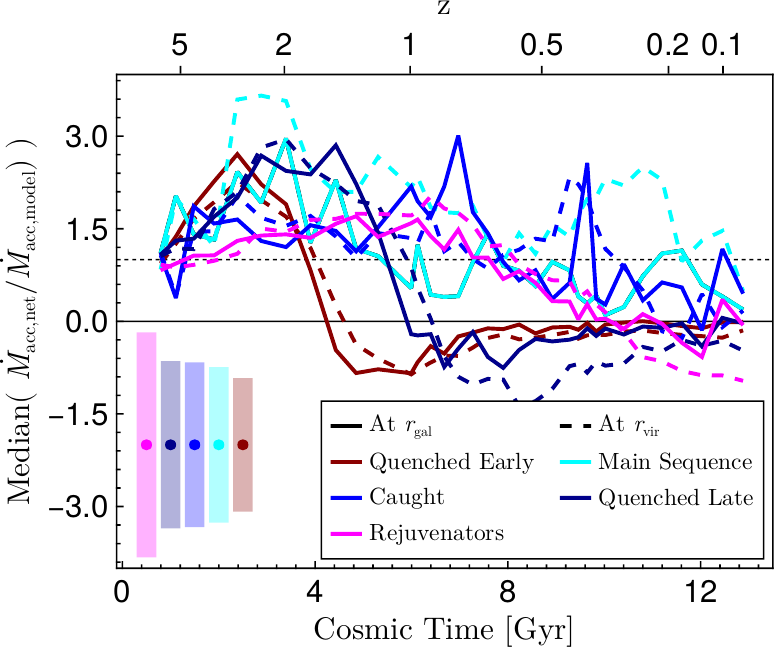}
    \caption{
	Median net gas accretion history evolution of each formation class across cosmic time. 
    We show the net accretion rates at the virial radii as dashed lines and as solid lines accretion rates onto the center of the halos at ten percent virial radius.
    Prior to combining the data, we normalized each accretion rate value by halo mass and redshift according to expectations from dark-matter assembly \citep[][]{dekel:2009}. 
    The horizontal dashed black line at a value of one highlights this expectation by definition.
    The median of main-sequence galaxies are in cyan, quenched in dark red, rejuvenating galaxies in magenta and caught galaxies in blue colors.
    The vertical bars outline a representative 50th-percentile scatter. 
    On average, accretion rates generally do not differ too much from the expectations. 
    Quenching galaxies collectively assume negative values correlating with the time of quenching of their star formation. 
    This correlation becomes evident from comparing the early- and late-quenching galaxy populations. 
    }
    {\label{fig:classes_accrates}}
    \end{center}
\end{figure}

We tested the link between gas flows and star formation by deriving the latter from flow rates and simplifying assumptions.
The median net accretion rate evolution used for the derivation is shown in \autoref{fig:classes_accrates}.
Specifically, we normalize each galaxy's net accretion rate by respective expected accretion rate \citep[][]{dekel:2009} and then calculate class median values at each snapshot across cosmic time.
When comparing the median net accretion rates onto the galaxies at $0.1\ R_\txt{vir}$ (solid lines) for to the values at the virial radii (dashed lines) for each class, the figure shows that net gas accretion evolves similar at both radii.
While main-sequence (cyan), caught (blue) and rejuvenator (magenta) type galaxies sustain positive net accretion rates roughly of the order of expectations from dark matter assembly \citep[][dashed vertical line]{dekel:2009}, quenching galaxies (dark red and dark blue) transition to even slightly negative values.
Notably, the main-sequence type galaxies (cyan) have consistently higher values at the virial radii than at the galaxy borders.

All galaxies tend to decline in net accretion towards lower redshifts. 
Quenched-type galaxies show the largest offset by having a clear peak at high redshifts and then dropping even below zero at lower redshifts.
Main-sequence types also peak early but never drop as much, and always stay among the highest accretion rates.
Rejuvenators peak later and less high. 
Caught-type galaxies maintain very high net accretion rates for a longer time compared to expectations from their halo masses.

Due to higher overall densities and velocities, the infall rates are higher in the centers of halos while all classes achieve lower and similar values at their virial radii.
The central inflow rates at ten percent of each virial radius of quenched-type galaxies drops severely off the total sample towards lower redshifts, which is consistent with their net accretion.

\section{Influence of AGN feedback}\label{sec:A_agnfeedback}

\begin{figure}
  \begin{center}
    \includegraphics[width=.9\columnwidth]{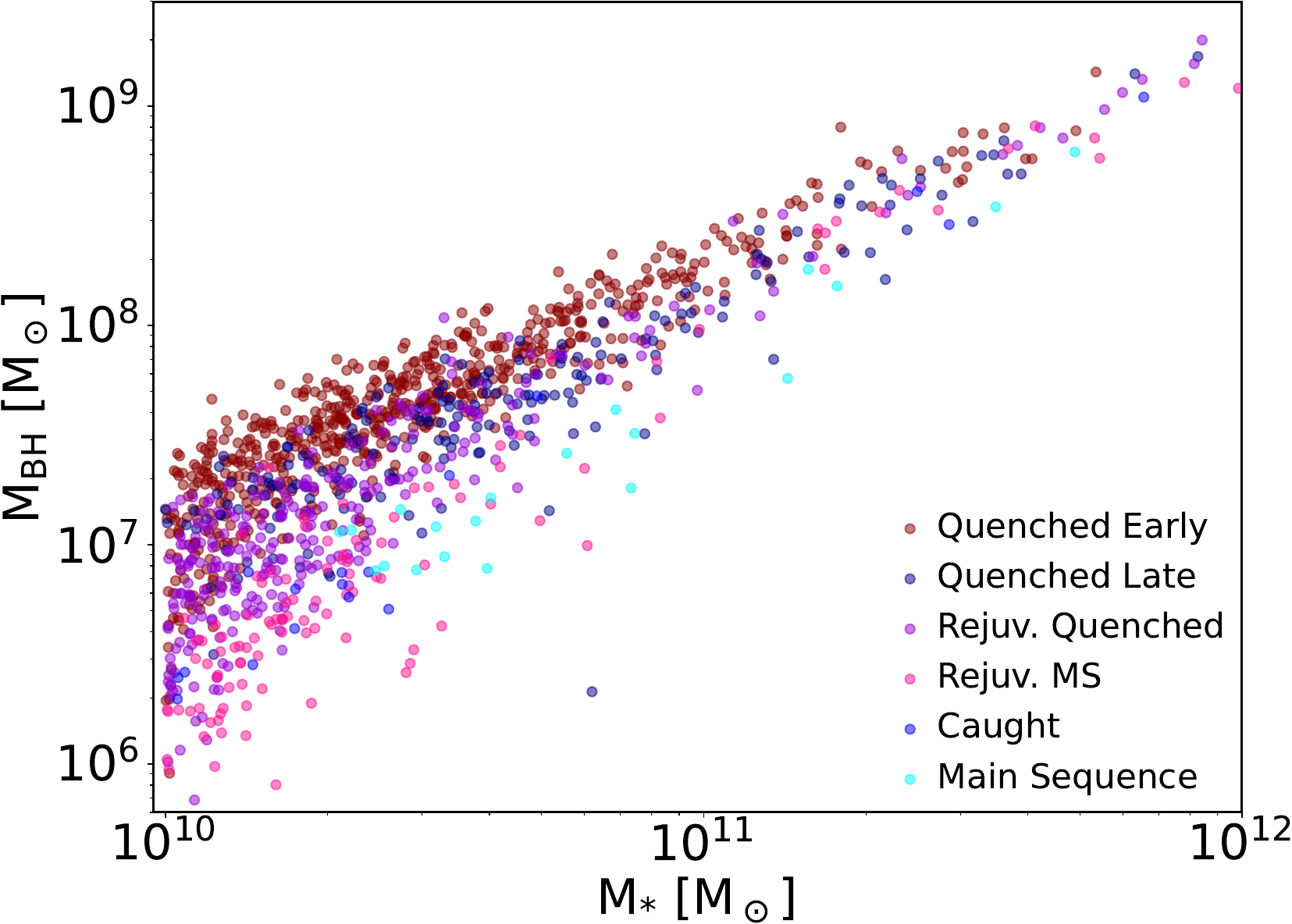}
    \caption{
        $\txt{M}_\txt{BH}$-$\txt{M}_*$ relation of the galaxy sample at $z\approx 0$. Especially for the mass range of $10^{10} \txt{M}_\odot \leq \txt{M}_\odot \leq 10^{11} \txt{M}_\odot $, a clear transition order from quenched galaxies along the Magorrian relation to lower black hole masses for star forming galaxies. Specifically, early-quenched galaxies (dark red) have the highest values, followed by late-quenched (dark blue), rejuvenated quenched (lilac), MS-evolved (cyan) and finally rejuvenating MS (pink) galaxies as well as caught-type galaxies (blue) at the lower black hole mass end.
    }
    {\label{fig:mbh_vs_mstar}}
    \end{center}
\end{figure}

\citet[][]{piotrowska:2022} have shown that AGN feedback quenching is strongly related to the mass of supermassive black holes in the centers of galaxies.
We therefore show in \autoref{fig:mbh_vs_mstar} the black hole masses and stellar masses of each galaxy at $z\approx 0$ colored according to its star formation evolution class to estimate the influence of AGN feedback on the evolution scenarios.
At any given stellar mass, quenched galaxies (dark red and dark blue circles) occupy the highest black hole masses.
This is likely due to how gas flows onto the central black hole, with high-z environments creating more isotropic flows \citep[][Seidel et al. in prep]{kimmig:2025a}, while galaxies which make a majority of their stars at later times or in on-off cycles do not as efficiently funnel the gas toward the center. This leads to lower black hole masses, with the most extreme case found for those evolving along the main sequence. We posit that galaxies which solely grew along the main sequence are the result of higher angular momentum gas that forms its stars in a disk without strongly feeding the central supermassive black hole.

\section{Environment with the $d_5$ parameter}\label{sec:A_enviroment}

\begin{figure}
  \begin{center}
    \includegraphics[width=.9\columnwidth]{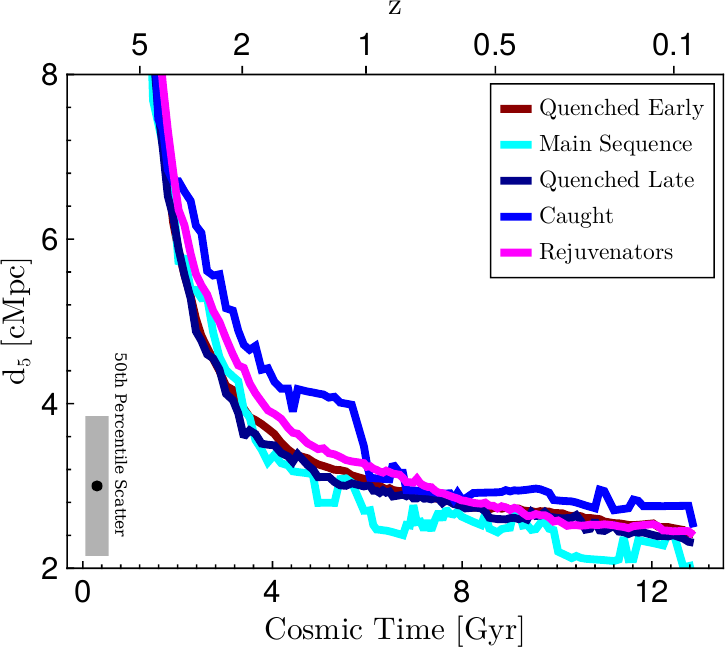}
    \caption{
        Evolution of the class median $d_5$ parameter values \citep[][]{remus:2023} across cosmic time. 
        Lower values indicate a denser environment of a galaxy.
        We chose comoving coordinates for a better comparison between across cosmic time.
        The 50th percentile scatter range shown in gray in the bottom left corner is similar for all classes and cosmic cosmic time.
        All classes show a very similar evolution with only faint differences with main-sequence types residing in denser environments consistently and caught-types found in less dense environments. 
    }
    {\label{fig:violin_d5}}
    \end{center}
\end{figure}

In \autoref{sec:environment}, we have discussed a clear correlation between gas inflow geometry at the halo outskirts and star formation rate evolution. 
While this shows that the environment plays an important role in the star formation evolution of a galaxy, the $d_5$ parameter is an unsuitable tracer of environment. 
We show in \autoref{fig:violin_d5} the evolution of median $d_5$ parameters for each class across cosmic time and find that, independent of redshift, the $d_5$ parameter barely varies between the classes. 
Only the smallest classes of main-sequence- and caught-type galaxies appear to reside in differently populated environments. 
Main-sequence galaxies tend to higher densities and caught-type galaxies to lower. 
This suggests that a denser environment neighboring galaxies only impacts the median specific star formation rate but not quenching or rejuvenation.
However, the differences are overall minuscule compared to the dominating overlap of violins.

\label{lastpage}
\end{document}